\documentclass[AMA,LATO1COL]{WileyNJD-v2}
\usepackage{siunitx}
\usepackage{rotating}
\newcommand{\myvec}[1]{\ensuremath\mathbf{#1}}
\newcommand{\myvecg}[1]{\ensuremath\boldsymbol#1}

\usepackage{tikz}
\usetikzlibrary{patterns}

\newcommand{\AC}{\raisebox{2pt}{\tikz{\draw[red,dotted,line width=0.9pt](0,0) -- (5mm,0);}}}
\newcommand{\AM}{\raisebox{2pt}{\tikz{\draw[red,dashed,line width=0.9pt](0,0) -- (5mm,0);}}}
\newcommand{\AF}{\raisebox{2pt}{\tikz{\draw[red,solid,line width=0.9pt](0,0) -- (5mm,0);}}}
\newcommand{\FC}{\raisebox{2pt}{\tikz{\draw[blue,dotted,line width=0.9pt](0,0) -- (5mm,0);}}}
\newcommand{\FM}{\raisebox{2pt}{\tikz{\draw[blue,dashed,line width=0.9pt](0,0) -- (5mm,0);}}}
\newcommand{\FF}{\raisebox{2pt}{\tikz{\draw[blue,solid,line width=0.9pt](0,0) -- (5mm,0);}}}
\newcommand{\Exp}{\raisebox{2pt}{\tikz{\filldraw (0,0) circle (2pt);}}}
\articletype{RESEARCH ARTICLE}

\received{21 October 2017}
\revised{01 June 2018}
\accepted{14 June 2018}

\begin{document}
\title{Mesh-adaptive simulations of horizontal-axis turbine arrays using the actuator line method}
\author{Georgios Deskos*}
\author{Matthew D. Piggott}
\authormark{DESKOS AND PIGGOTT}
\address{Department of Earth Science and Engineering, South Kensington Campus, Imperial College London, SW7 2AZ, UK}
\corres{*G. Deskos, Department of Earth Science and Engineering, South Kensington Campus, Imperial College London, SW7 2AZ, UK. \\
E-mail : g.deskos14@imperial.ac.uk}
\fundingInfo{Engineering and Physical Sciences Research Council (EPSRC), 
Grant Numbers: EP/R007470/1 and EP /L000407/1.}

\abstract[Abstract]{
Numerical models of the flow and wakes due to turbines operating within a real-scale offshore wind farm can lead to a prohibitively large computational cost, particularly when considering blade-resolved simulations. With the introduction of turbine parametrisations such as the actuator disk (ADM) or the actuator line (ALM) models this problem has been partially addressed, yet the computational cost associated with these simulations remains high. In this work we present an implementation and validation of an ALM within the mesh-adaptive 3D fluid dynamics solver, Fluidity, under a uRANS-based turbulence modelling approach. A key feature of this implementation is the use of mesh optimization techniques which allow for the automatic refinement or coarsening of the mesh locally according to the resolution needed by the fluid flow solver. The model is first validated against experimental data from wind tunnel tests. Finally, we demonstrate the benefits of mesh-adaptivity by considering flow past the Lillgrund offshore wind farm.}

\keywords{actuator line model; mesh optimization; uRANS turbulence models; Lillgrund offshore wind farm}
\jnlcitation{\cname{%
\author{Deskos G.}, and
\author{M. D. Piggott}} (\cyear{2018}),
\ctitle{Mesh-adaptive simulations of horizontal-axis turbine arrays using the actuator line method}, \cjournal{Wind Energy},
\cvol{2018;00:1--6}.}

\maketitle
\section{Introduction}\label{sec:Intro}
Turbine parametrisation models (TPMs) such as the actuator line model (ALM) and the actuator disk model (ADM) exhibit a large number of advantages compared to blade--resolved simulations, both in terms of their respective computational efficiency but also as far as their implementation within a CFD solver is concerned. First, by using TPMs the number of degrees of freedom needed by the fluid solver is significantly reduced since the boundary layer of the individual blades is no longer required to be resolved. Second, the introduction of the momentum source to represent the motion of the blades circumvents the need to use either a rotating or an overlapping mesh strategy to capture the motion of the turbine rotor. These two factors have rendered the use of TPMs a computationally affordable alternative approach for the modelling of large-scale wind farms. Hence, the ADM was used by \cite{AmmaraEtAl2002,JimenezEtAl2007,CalafEtAl2010,NilssonEtAl2015,WuPorteAgel2015} to model the wake field and predict the power output of operating offshore wind farms (e.g. Lillgrund, Horns Rev) while \cite{LuPorte-Agel2011,ChurchfieldEtAl2012,ChurchfieldEtAl2012a} undertook ALM simulations to solve for the wake field as well as to obtain statistics for blade loads. In all of these studies either a uniform mesh or a block-mesh strategy was employed within the presented simulations. Such an approach requires a priori knowledge of the wake length and width, or otherwise a large volume of the computational domain to be a assigned as the refined ``wake region''. Moreover, as more realistic simulations are required for utility--scale wind farms, including changes in the wind (and therefore wake) direction, the ``refined wake regions'' will need to be expanded to a greater extend in order to provide the required resolution of such simulations. Undoubtedly, this approach is not optimal for the discretization of the domain, as even a moderate expansion (in the order of a few decades of metres) of the wake region can significantly increase the number of degrees of freedom (DoF). Inherently, some sort of flexible mesh adaptivity procedure (e.g. the dynamic mesh optimization approach used here) which is employed during the course of the simulation would be an attractive approach to consider for the above described problem. This particular need has already been expressed in previous studies. For instance, Churchfield et al. \cite{ChurchfieldEtAl2012a} mentions that \textit{``\ldots adaptive mesh refinement would be useful in providing higher resolution only where necessary, but may incur a run-time penalty in performing the refinement and the processor load balance \ldots}''. In a similar note, Nilsson et al. \cite{NilssonEtAl2015} also used different resolution grids and pointed out that the simulations using the most refined grids were abandoned due to limitations on the available computational resource.

Combining turbine parametrisation models with mesh--optimisation techniques was first considered by \cite{CreechEtAl2011} and more recent implementations can also be found in \cite{AbolghasemiEtAl2016} and \cite{CreechEtAl2017}, each employing a different turbine parametrisation model (actuator volume and actuator disk and actuator line based, respectively). 
Mesh--adaptivity has been used in conjunction with blade--resolved simulations by \cite{PotsdamMavriplis2009,WissinkEtAl2010a,WissinkEtAl2010b} who employed an unstructured overlapping mesh strategy to obtain the near--body solution and either unstructured or Cartesian mesh--adaptive solvers for the off--body flow. Such dual mesh approaches have been effectively used to calculate wind turbine and rotorcraft wakes and it has been postulated that the computational efficiency of the approach is thanks to the off--body dynamic mesh--adaptivity solver. Indeed, a significant amount of computational resources can be saved by using an optimal local refinement or coarsening of the mesh while at the same time maintaining the desired levels of solution accuracy. In the context of large--scale wind farm simulations, mesh--adaptivity was used by \cite{CreechEtAl2015} and \cite{KirbyEtAl2017} to simulate the wakes developed within the Lillgrund offshore wind farm. Both studies demonstrated the ability of a mesh--adaptive solver (either unstructured or Cartesian--based) to be coupled with turbine parametrisation models or overlapping mesh strategies, respectively, and to be used as a high--fidelity, multiscale wind farm modelling tool. 

In the above mentioned mesh--optimization/adaptivity algorithms and studies the obtained solutions have not been compared with traditional static mesh solutions using the same solver in order to provide a rigorous estimate of either the potential accuracy gains and/or reductions in computational effort. Such questions are important in shedding light on the efficiency and accuracy of mesh--adaptive solvers, and more specifically on their applicability to wind energy research. To partially address these questions, we present herein the implementation and validation of an ALM which employs dynamic mesh optimization techniques. The optimization of the mesh is achieved through a strategy which allows control over both the numerical error and mesh size at run time \cite{PiggottEtAl2005,PiggottEtAl2008,PiggottEtAl2009}. Both the ALM and mesh--adaptivity approach are developed within the open source code Fluidity \cite{PiggottEtAl2008,Fluidity.Manual.v.4.1.12} which is a general purpose unstructured mesh based finite element solver. In addition, the fluid flow is modelled in this work using an unsteady Reynolds-averaged Navier-Stokes (uRANS) based approach, combined with the $k$--$\omega$ SST turbulence model \cite{Menter1994}. Before proceeding to the large-scale simulations, the accuracy of the new ALM implementation in Fluidity (combined with the uRANS configuration) is investigated through comparisons with a series of wind tunnel tests \cite{KrogstadEriksen2013,PierellaEtAl2014} for the power and wake of one and two turbines operating in line. The model shows very good agreement with the rotor's thrust and power coefficients predictions as well as the wake field. These comparisons give us confidence that the model can predict the wake characteristics with high accuracy when real-world scale wind farms are considered. To demonstrate the efficiency of the mesh optimization approach we compare the results from an adaptive-mesh simulation with those from a static pre-refined mesh simulation for the Lillgrund offshore wind farm and data from the literature \cite{Dahlberg2009}.

The paper starts with section \ref{sec:Theory} which introduces the mesh-adaptive fluid solver and section \ref{sec:ALM}  which discusses the implementation of the turbine parametrisation. In Section \ref{sec:ValVer}, the two mesh approaches (fixed vs. adaptive) are validated against the experimental data of \cite{KrogstadEriksen2013,PierellaEtAl2014} while in section \ref{sec:LillgrundSimulation} simulation of the Lillgrund offshore wind farm are undertaken. The mesh optimization techniques are presented from the point of view of the same solver (Fluidity) and its contribution to increasing accuracy and reducing computational cost are finally discussed thereafter in section \ref{sec:Discussion}.
\section{Model implementation}\label{sec:Theory} 
\subsection{Unsteady RANS formulation of the governing equations}\label{sec:LESform} 
The wind turbine wakes are modelled using the unsteady Reynolds-averaged Navier-Stokes (uRANS) equations, in which the velocity is decomposed into mean and fluctuating (turbulent) components. Within uRANS the governing equations take the form
\begin{equation}
\nabla \cdot  \myvec{\overline{u}} = 0, 
\label{eq:rans_continuity}
\end{equation}
\begin{equation}\label{eq:rans_momentum}
\rho\frac{\partial \myvec{\overline{u}}}{\partial t}
+\rho \myvec{\overline{u}} \cdot \nabla \myvec{\overline{u}}
= -\nabla \overline{p}
+ \mu \nabla^2 \myvec{\overline{u}}
- \nabla \cdot (\overline{\rho \myvec{u'} \otimes \myvec{u'}})
+ \myvec{F}_T,
\end{equation}
where $\myvec{\overline{u}}$ is the time-averaged component of velocity, $\myvec{u'}$ is the fluctuating velocity component, $\overline{p}$ is the mean pressure, $\mu$ is the dynamic 
viscosity of the fluid, and $\myvec{F}_T$ is a momentum source term computed at each time step from the ALM. The term $-\nabla \cdot (\overline{\rho \myvec{u'} \otimes \myvec{u'}})$ 
is a residual term from the application of the time-averaging operator on the outer product of the fluctuating velocity components, called the Reynolds stress tensor 
$\overline{\overline{\myvecg{\tau}}}_R$. The presence of the Reynolds stress tensor in \eqref{eq:rans_momentum} introduces a number of additional 
unknowns and therefore the Boussinesq approximation is adopted to provide closure to the system of equations. That is, the Reynolds stresses are related to the time-averaged turbulence 
kinetic energy $k=\tfrac{1}{2}\overline{\myvec{u'}\cdot\myvec{u'}}$ and strain-rate tenso r via
\begin{equation}\label{eq:Boussinesq}
- \nabla \cdot (\overline{\rho \myvec{u'} \otimes \myvec{u'}})
= \overline{\overline{\myvecg{\tau}}}_R
= -\frac{2}{3}k\rho \myvec{I} + \mu_T \bigg(\nabla\myvec{\overline{u}}+
(\nabla\myvec{\overline{u}})^T \bigg).
\end{equation}
where $\mu_T$ is the eddy viscosity and $\myvec{I}$ the unit tensor. To compute $k$ we make 
use of the standard $k-\omega$ SST model proposed by Menter \cite{Menter1994} which requires the solution of two additional transport equations:
\begin{equation}\label{eq:k-omegaSST_k}
     \rho \frac{\partial k}{\partial t} 
     + \rho \myvec{u} \cdot \nabla k 
     =
     \nabla \cdot \bigg(\big(\mu + \mu_T\sigma_k \big) \nabla k \bigg)
     + \tilde{P}_k 
     - \rho \beta^* k \omega, 
\end{equation}
and
\begin{equation}\label{eq:k-omegaSST_epsilon}
     \rho \frac{\partial \omega}{\partial t} 
     + \rho \myvec{u} \cdot \nabla \omega 
     =
      s
      ./b\nabla \cdot \bigg(\big(\mu + \mu_T\sigma_{\omega} \big)\nabla \omega  \bigg)+
     \bigg(\frac{\rho \alpha}{\mu_T} \bigg) \tilde{P}_k 
     - \rho \beta \omega^2
     +2 \bigg( 1- F_1 \bigg) \rho \sigma_{\omega_2} \frac{1}{\omega} \nabla k \nabla\omega,
\end{equation}
where
\begin{equation}\label{eq:eddy_viscosity}
     	\mu_T =\rho \frac{k}{\omega}, 
\end{equation}
is the eddy viscosity and $F_1$ a blending function defined in \cite{Menter1994}. The tildered quantity $\tilde{P}_k$ 
denotes a limiting turbulence kinetic energy production given by
\begin{equation}\label{eq:limiting_prod}
\tilde{P}_k=\min(P_k,10 \rho \beta^* \omega),
\end{equation}
which is applied to prevent the build-up of turbulent energy in stagnation regions 
\cite{MenterEtAl2003}. Finally, the closure coefficients 
$\sigma_k$, $\sigma_{\omega}$, $\alpha$, $\beta$ and $\beta^*$ 
are selected by linear interpolation using the blending function value $F_1$. Further 
information for the model implementation can be found in the appendix of 
\cite{AbolghasemiEtAl2016} and the references therein.
\subsection{Numerical implementation and mesh optimization}\label{subsec:NumImplem}
The system of governing equations, including the additional scalar transport 
equations required for the turbulence modelling, have been discretized within the 
open-source code Fluidity \cite{Fluidity.Manual.v.4.1.12}. Fluidity is a general purpose 
three-dimensional, unstructured mesh, finite element/control volume based PDE solver 
\cite{PainEtAl2005,PiggottEtAl2005,PiggottEtAl2009} with the ability to make use of 
optimization based anisotropic mesh adaptivity. 
For our analysis, the continuity and momentum equations are 
discretized using mixed finite elements for which piecewise-linear discontinuous basis functions 
are used to represent velocity, while continuous piecewise-quadratic basis functions are used for pressure over tetrahedral elements (the so-called P1$_{\textnormal{DG}}$--P2 element pair). 
This scheme is known to be a Ladyzhenskaya-Babuska-Brezzi (LBB) stable combination \citep{DoneaHuerta2005} and to perform well for 
advection-dominated problems \cite{CotterEtAl2009}. The formulation also uses a slope limiter 
\cite{Kuzmin2010} to ensure a robust solution for the velocity and pressure fields. 
The $k$--$\omega$ SST model makes use of a control volume based 
discretization \cite{PiggottEtAl2009} with flux limiters to help prevent oscillatory behaviour 
of the turbulent kinetic energy $k$.  For time marching the second-order 
accurate Crank-Nicolson scheme is used and is combined with an additional explicit sub-cycling 
approach for momentum advection \cite{Fluidity.Manual.v.4.1.12}.	 	

The underlying unstructured tetrahedral mesh is also subject to optimization-based 
adaptivity algorithms which are used in order to improve the quality of the mesh as well 
as provide higher or lower resolution at locations which are identified by the solver.
For instance, the introduction of the ALM momentum source will create a 
requirement for a particular edge length over the assigned rotor volume. This is achieved here 
through the specification of a scalar field which identifies the region that the rotating actuator lines will 
occupy during the simulation. At the same time two additional fluid properties, the velocity 
vector field and the turbulence kinetic energy (a scalar field) are also used to guide
the mesh-optimization process. This is achieved through the derivation of a metric tensor field. If we consider a single scalar field $\phi$ which we want to adapt our mesh to optimally resolve, we form a metric tensor, $\mathcal{M}_{\phi}$, by first computing the Hessian, $H_{\phi}$, of that scalar field and defining 
\begin{equation}
\mathcal{M}_{\phi}=\frac{1}{\epsilon_{\phi}} |\bar{H}_{\phi}|,
\end{equation}
where $\epsilon_{\phi}$ is a user-defined weight for field $\phi$ ($\epsilon_{\phi}$ can in some sense be considered a requested error -- a smaller value leading to a larger $\mathcal{M}_{\phi}$ and consequently a finer mesh). The Hessian encodes information about the curvature of the scalar field and includes both spatial and direction information; we desire finer mesh resolution at location and in directions where curvature is high, and coarser resolution where the solution is close to linear. $|\bar{H}_{\phi}|$ indicates that we are interested in the magnitudes of curvatures when deciding on optimal mesh resolution, and do not care about the sign.
Once metric tensor fields have been calculated for all the scalar fields (for the velocity vector we consider each component separately) we wish to adapt to, a final metric is obtained through 
superimposition of the individual metrics \cite{PainEtAl2001}. 
At this stage additional constraints on the total number of elements in the calculation,
and/or maximum and minimum edge length, and the maximum rate at which edge lengths can vary
in space can all be incorporated.  The metric can then be used to measure the length of vectors -- primarily the edge lengths of elements.  A perfectly optimized mesh in physical space is defined as one which is made up of unit length edges in metric space. The inhomogeneous and directionally dependent nature of the Hessians and hence the metrics thus leading to a mesh which is variable in both space and direction (i.e. is potentially anisotropic). An optimization functional is defined which measures how well the current mesh achieves this goal, and a series of topological 
operations are performed on the current mesh to improve this agreement. These operations include 
edge collapse and splitting, and face to face and face to edge swapping \cite{PainEtAl2001,PiggottEtAl2009}. 
Finally, conservative mesh to mesh interpolation is used to transfer solution data from
the old to the new mesh \cite{FarrellEtAl2009,FarrellMaddison2011}. The entire mesh optimization process is conducted every user-defined number of time period, termed the adaptation period $T_{\textnormal{adapt}}$. 
Further information on the mesh optimization process, including its parallelisation, may be found in \cite{PainEtAl2001,PiggottEtAl2005,PiggottEtAl2008,PiggottEtAl2008b,PiggottEtAl2009} and the references therein. 
\section{Turbine parametrisation}\label{sec:ALM}
Our actuator line model (ALM) follows the approach of \cite{SorensenShen2002,TroldborgEtAl2009}
in which the turbine blades are represented by rotating virtual lines -- the actuator lines (AL). 
Point forces are computed along each AL, (at each blade element's midpoint) using the relative 
velocity $\myvec{U}$ extracted from the fluid solver by evaluating the globally-defined finite element 
solution at these points, the solid body velocity $\myvec{U}_b$ of each point, and the lift and drag coefficients obtained from 
look-up tables using airfoil data for the blade element's respective profile. 
Extra care has been taken for the elements near the blade tip for which the tip loss correction model of \cite{ShenEtAl2005} is used.
The tower behind the turbine is also incorporated by adopting a model similar to \cite{SarlakEtAl2015} -- the tower is represented by an 
actuator line with element having a cylindrical cross-section of constant drag coefficient $C_D=1$ and a time--dependent lift coefficient 
$C_L$ which is ``tuned'' in order to reproduce the Von K{\'a}rm{\'a}n street behind the cylinder,
\begin{equation}
C_L= A\sin (2 \pi f t),
\label{eq:tower_lift}
\end{equation}
where $A$ is the amplitude of the ``dynamic lift force'' and $f=0.2 \times 
U_{\infty}/D_{\textnormal{tower}}$ 
is the Strouhal number, based on the uniform velocity and the diameter of the tower 
$D_{\textnormal{tower}}$. For all simulations presented in this work the hub and nacelle of the turbines
are not modelled.
The final ALM forces are projected onto the fluid mesh and represented in the governing equations via the momentum source term $\myvec{F}_T$. 
To ensure a smooth transition from a concentrated point AL force $\myvec{f}_{AL}$ to the source term $\myvec{F}_T$, a smoothing interpolation function 
is used, 
\begin{equation}
\myvec{F}_T =-\frac{1}{\epsilon^3 \pi^{3/2}} \exp \bigg(-\frac{|\myvec{r}|^2}{\epsilon^2}\bigg) \myvec{f}_{AL}.
\label{eq:smoothing_pointkernel}
\end{equation}
where $|\myvec{r}|$ is the distance of the mesh point from the AL node and $\epsilon$ is 
a smoothing parameter selected after taking into account the mesh size, drag force $C_D$, chord size $c$
and $V_{\textnormal{elem}}$ is the volume of the element in which the AL node lies,
\begin{equation}
\epsilon = \max\bigg[\frac{c}{4},4\sqrt[3]{V_{\textnormal{elem}}},\frac{c C_D}{2} \bigg],\label{eq:smoothing_parameter}
\end{equation}
consistent with the recommendations of \citep{Martinez-TossasEtAl2017}.
\section{Model validation}\label{sec:ValVer}
\begin{figure}[t]
\centering
\includegraphics[width=0.7\linewidth]{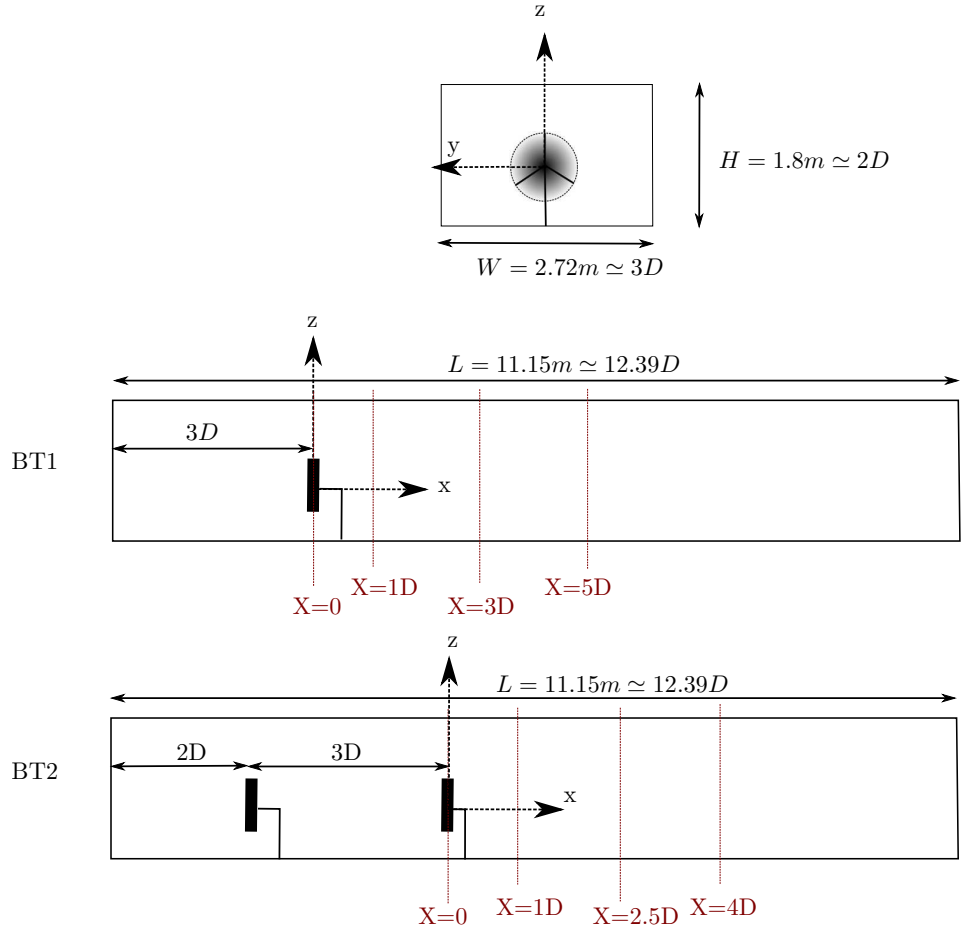} 
\caption{Schematic representation from the two ``Blind Tests'' set ups.}
\label{fig:BlindTests_scematic}
\end{figure}

The newly implemented ALM is validated using data from a series of ``Blind Test'' workshops organized by NoWiTech and NoCOWE and which was obtained from the wind tunnel 
facilities of NTNU \cite{KrogstadEriksen2013,PierellaEtAl2014}. We will refer to the two papers that reported the data as ``blind tests'' or when 
mentioned individually as ``blind test 1'' (BT1) and ``blind test 2'' (BT2). The wind tunnel facility used for the two blind tests is \SI{11.15}{\meter} long, 
\SI{2.72}{\meter} wide, and \SI{1.8}{\meter} high. In (BT1) a single turbine with rotor diameter $D=\,$\SI{0.894}{\meter}, hub height equal to 
$H_{\textnormal{hub}}=\,$\SI{0.817}{\meter} is placed at a distance of $2D$ from the wind tunnel inlet. The turbine has three blades consisting of \num{14}$\%$ NREL S826 airfoils, 
a tower (support structure) with diameter $D_{tower}=\,$\SI{0.11}{\meter} and is designed to operate optimally for a tip speed ratio of $\lambda=\,$\num{6}. For BT1, power and 
thrust coefficients are reported by \cite{KrogstadEriksen2013} for a large range of tip speed ratios (\numrange{5}{11.5}) while wake statistics including the stream-wise 
velocity deficit and the turbulence kinetic energy for $\lambda=6$. On the other hand, BT2 involves wake predictions from two similar turbines (with slightly different 
diameters $D_1=\,$\SI{0.944}{\meter} and $D_2=\,$\SI{0.894}{\meter} where the subscripts 1 and 2 correspond to the ``Front'' and the ``Rear'' turbines respectively) operating in-
line. The front turbine is located at $2 D_2=\,$\SI{1.788}{\meter} from the tunnel's entrance while the rear is placed at a distance $3 D_2=\,$\SI{2.686}{\meter} from the 
first as shown in figure \ref{fig:BlindTests_scematic}. BT2 reports three operational scenarios, in which the front turbine operates with $\lambda_1=\Omega 
R/U_{\infty}=\,$\num{6}, and the rear with $\lambda_2=\,$ \numlist{4;7;2.5}. In both BT1 and BT2 the mean free--stream velocity was measured to be \SI{10}{\meter \per \second} 
whereas the vertical profile of the ambient turbulence intensity found to be nearly uniform with a value of $I\approx\,$\num{0.3}$\%$. For our computations, we also assume a 
uniform velocity and TKE profiles with initial and inlet conditions $U=\,$\SI{10}{\meter \per \second} and $k=\,$\SI{1.3e-3}{\meter^2 \per \second^2} 
respectively while the turbulent frequency is also considered to be uniform and equal to $\omega=\,$\SI{0.5}{\per \second}. Finally, the tower model is enabled for all the wind 
tunnel simulations with the lift coefficient given a value of $A=\,$\num{0.3}, consistent with the study of \cite{SarlakEtAl2015}.
\subsection{Mesh convergence study}\label{sec:verif} 
\begin{figure}[!htb]
\begin{center}
\includegraphics[width=0.4\linewidth]{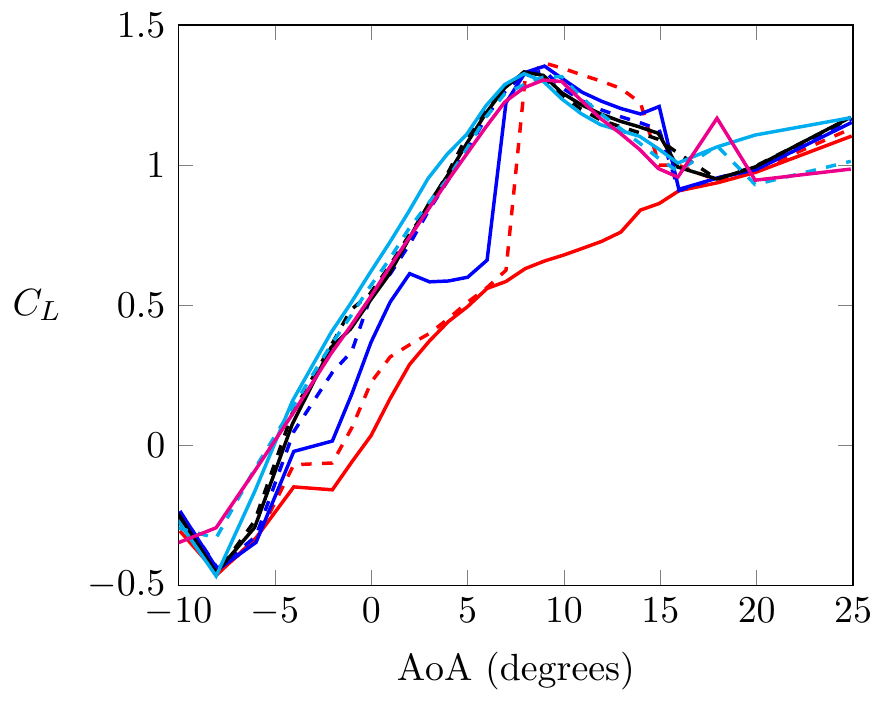} 
\includegraphics[width=0.4\linewidth]{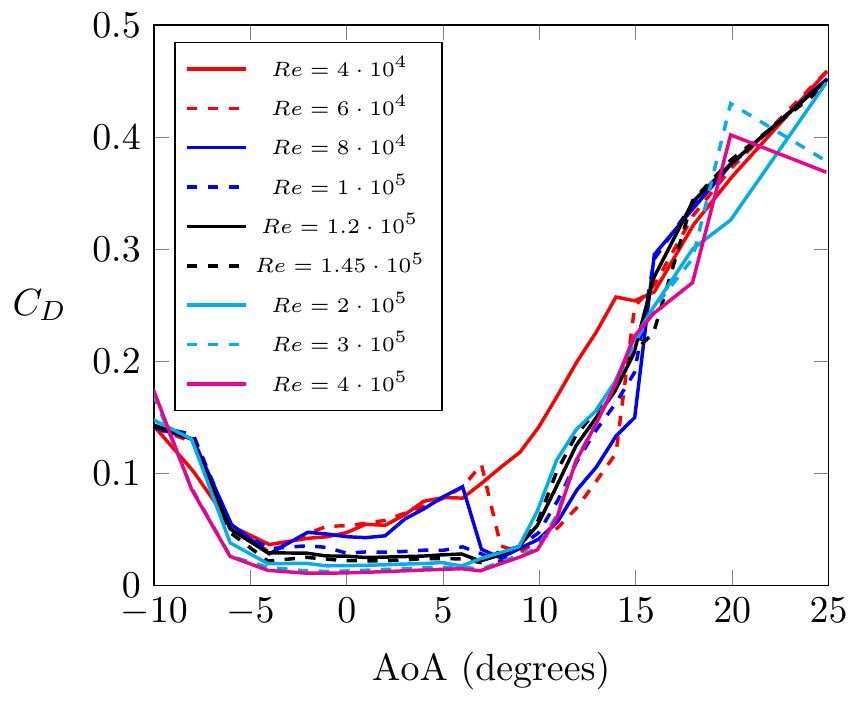}
\end{center}
\caption{Lift and drag coefficients as a function of the angle of attack (AoA) \cite{Sarlak2014} used for the model validation.}
\label{fig:lift_drag_coeffs}
\end{figure}

\begin{figure}[t]
\centering
\includegraphics[width=0.47\linewidth]{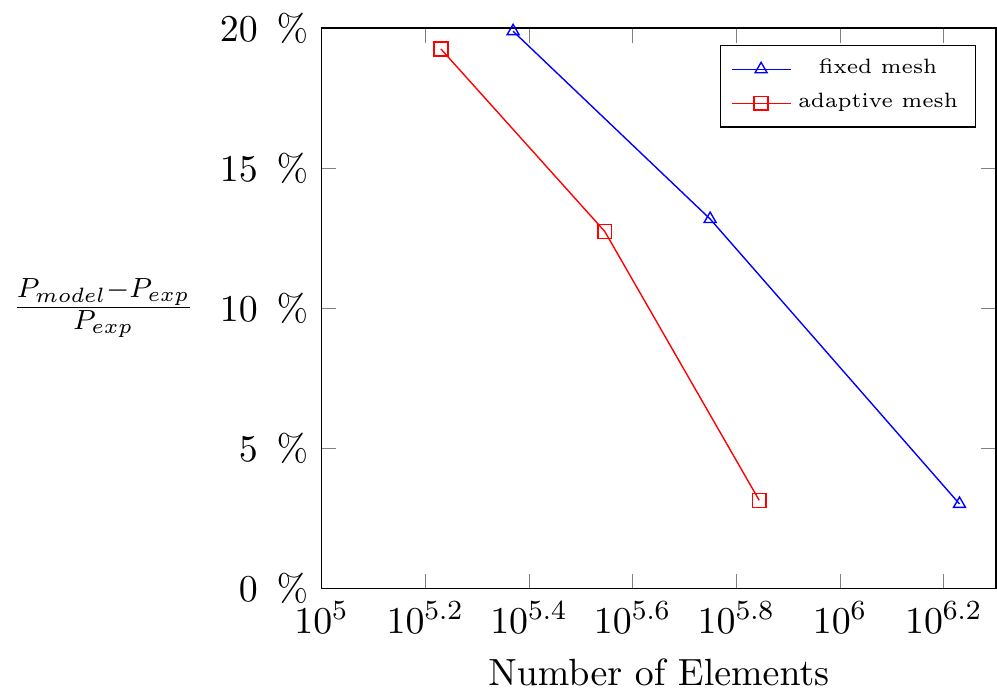}
~
\includegraphics[width=0.40\linewidth]{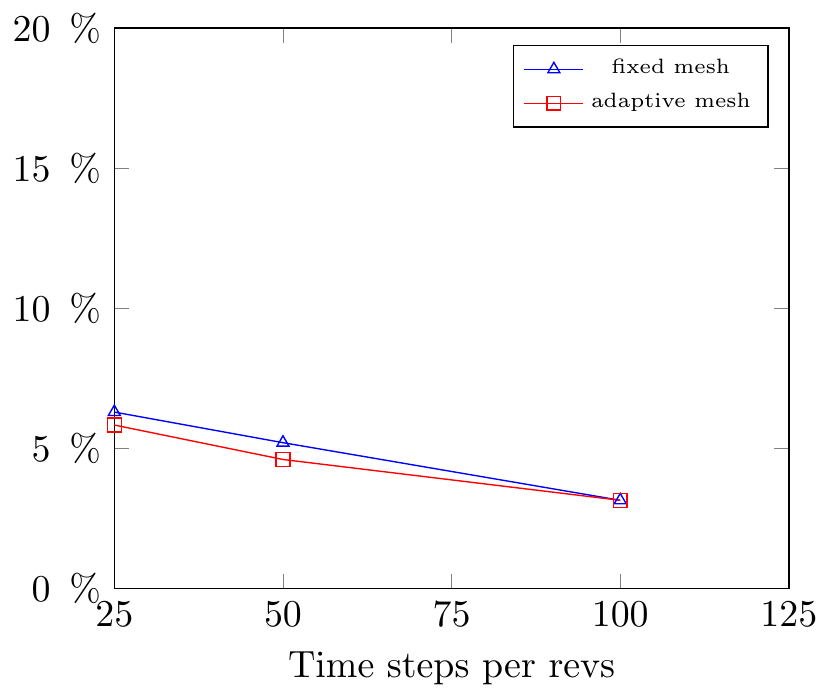}
\caption{Convergence study with respect to spatial and temporal resolution.}
\label{fig:Conv}    
\end{figure}

The design power coefficient from the single turbine experiments (BT1) is used as a representative quantity for our mesh size and time step convergence studies. For the mesh convergence study 
we consider six set-up cases by assuming two mesh types (uniform fixed, adaptive) and three minimum element edge lengths $h=\,$\SIlist{0.1;0.075;0.05}{\meter} as shown in table 
\ref{table:cases}, while the same time step $\Delta t=\,$\SI{0.005}{\second} is used for all simulations. We should note here that the adaptive simulations are also restricted by a maximum 
edge length of \SI{0.5}{\metre}. Additionally, all adaptive simulations were conducted using an adaptation period $T_{\textnormal{adapt}}=\,$\SI{2.5}{\second}. On the other hand, the time step 
convergence study considers three time step sizes $\Delta t =\,$\SIlist{0.02;0.01;0.005}{\second}, corresponding to approximately \numlist{25;50;100} time steps per rotor revolution, and 
using the finest mesh ($h=\,$\SI{0.05}{\meter}) for both the fixed and adaptive mesh.
\begin{table}[t]
\centering
\begin{center}
\captionof{table}{Table of simulation set-up cases.}\label{table:cases}
\begin{tabular}{cccccc}
\hline
Case  & Min. Mesh size h [m] & Average. Num. Elem. & Mest Type  & Tag\\
\hline
01    & 0.1     & 234,098      &  Fixed       & FC\\
02    & 0.075   & 561,611      &  Fixed       & FM\\  
03    & 0.05    & 1,699,635    &  Fixed       & FF\\ 
04    & 0.1     & 169,939      &  Adaptive    & AC\\ 
05    & 0.075   & 351,688      &  Adaptive    & AM\\ 
06    & 0.05    & 698,236     &  Adaptive    & AF\\ 
\hline
\end{tabular}
\end{center}
\end{table}
The lift and drag coefficients for the $14\%$ NREL S826 section are taken from 
\cite{Sarlak2014} and include data spanning a Reynolds number from 
\numrange{4e4}{4e6} and angles of attacks from \numrange{-10}{25} as shown in figure 
\ref{fig:lift_drag_coeffs}.
The turbine performance and thrust coefficients are computed based on the uniform 
upstream velocity $U_{\infty}$ and the nominal radius $R=D/2$ via
\begin{equation}
C_T= \frac{T}{\tfrac{1}{2} \rho \pi R^2 U_{\infty}^2},
\label{eq:Thrust_Coeff}
\end{equation}
\begin{equation}
C_P= \frac{P}{\tfrac{1}{2} \rho \pi R^2 U_{\infty}^3}.
\label{eq:Power_Coeff}
\end{equation}
To obtain mean values for the power and thrust coefficients we run long enough simulations  and allow the turbines to undergo more than 100 revolutions. In figure 
\ref{fig:Conv} the relative error in the predicted power coefficient is plotted against the average number of elements used in the simulations. We observe that for a 
given minimum element edge length, the two approaches (fixed and adaptive mesh) yield very similar relative errors. However, the adaptive mesh method required a smaller number of elements due 
to the optimum and flexible use of the underlying mesh within the computational domain. Therefore, within the ALM/uRANS configuration considered here, the minimum 
edge length in the mesh can be considered the primary factor determining the accuracy of the model, while adaptivity is used primarily to reduce the overall number of degrees of freedom, and 
therefore the associated computational cost. Later we will further demonstrate the potential advantages of mesh--adaptivity by undertaking large-scale wind farm simulations. In that case the 
substantially fewer elements required during the spin-up period of the wake development leads to an important saving in CPU time. On the other hand, the temporal convergence study (right-hand 
side of figure \ref{fig:Conv}) shows that as we reduce the time step (and thus increase the number of time steps used during one rotor revolution), the accuracy of the model converges to the 
element-based maximum obtainable accuracy. Again, the plotted figures correspond to simulations using the minimum edge length $h=\,$\SI{0.05}{\meter}. Therefore, the element edge length and 
the time step used in all simulations presented hereafter, will be based upon this preliminary convergence study. We should also note that once we have selected an element edge length for 
our analysis, the magnitude requirement on the time step is entirely dictated by the ALM and not the stability of the fluid solver, which allows for a far more relaxed 
condition assuming a Courant number of near unity.
Lastly, the smoothing parameter $\epsilon$ varies with the element's edge length and for all simulations hereafter 
is taken equal to \num{2.5} times the edge length, which also satisfies equation \eqref{eq:smoothing_parameter}.
\subsection{Wind Tunnel Tests}\label{subsec:BlindTests}
\begin{figure}[t]
\centering
\includegraphics[width=0.3\linewidth]{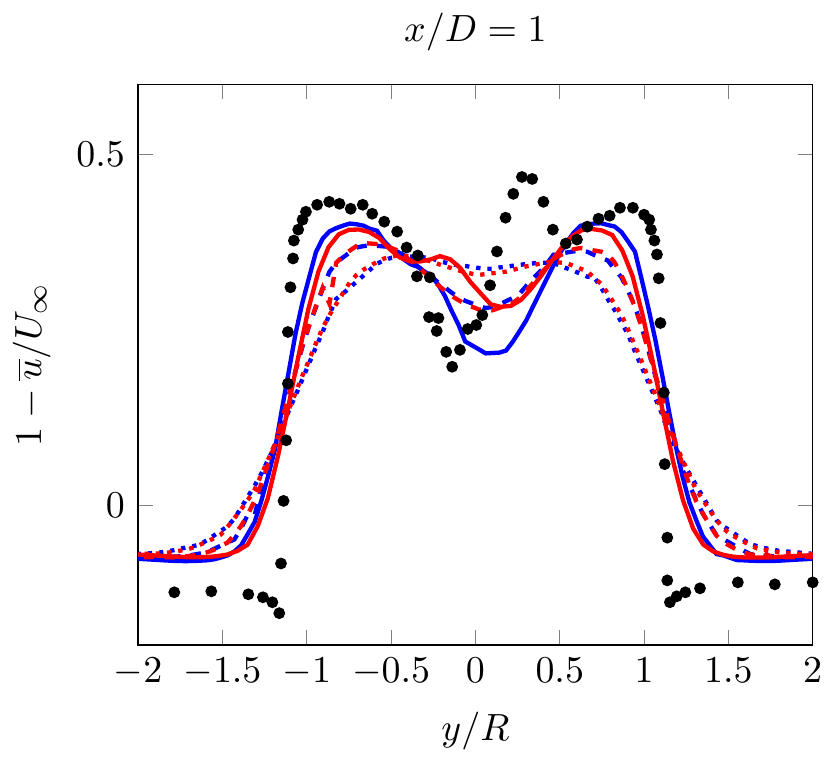} 
~
\includegraphics[width=0.3\linewidth]{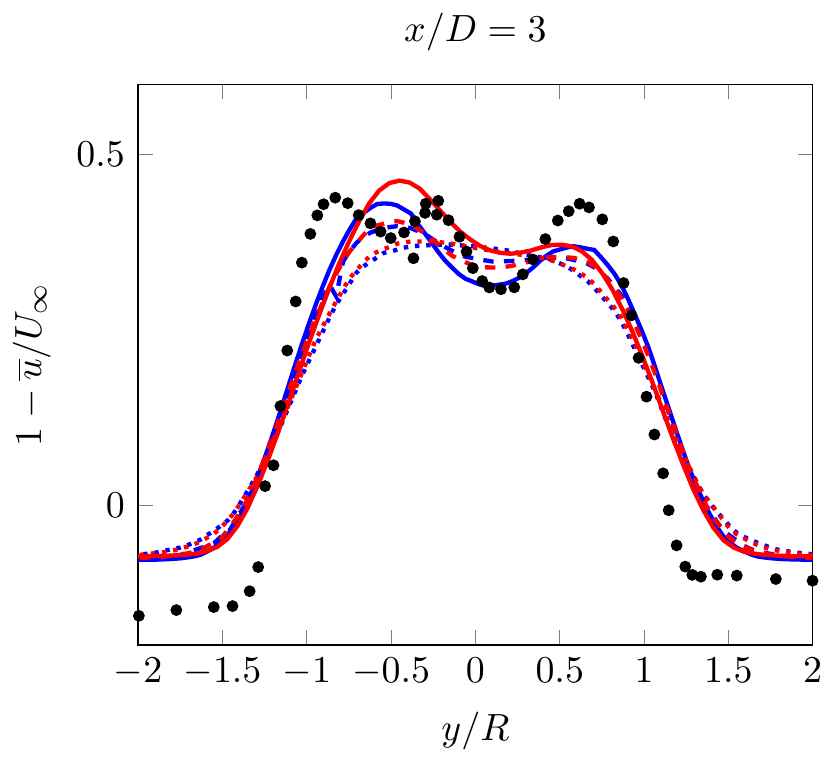} 
~
\includegraphics[width=0.3\linewidth]{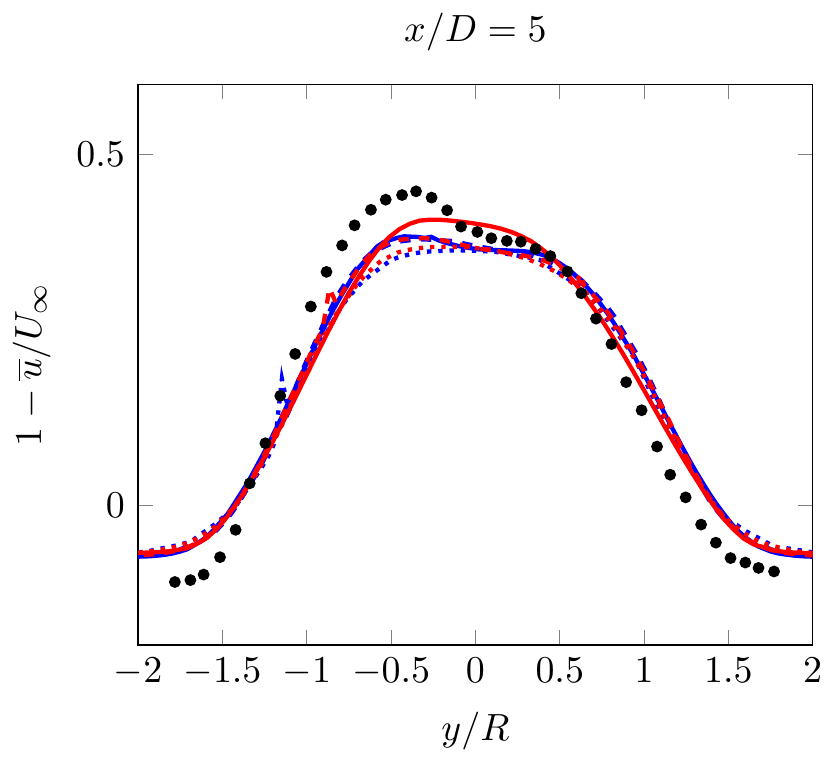} 
\caption{Blind Test 1: Horizontal mean stream-wise velocity profiles at 
$x/D=\,$\numlist{1;3;5}. The plotted lines correspond to: fixed coarse (FC) 
(\protect\FC), fixed medium (FM) (\protect\FM), fixed fine (FF) (\protect\FF), adaptive 
coarse (AC) (\protect\AC), adaptive medium (AM) (\protect\AM), adaptive fine (AF) 
(\protect\AF) and the symbols (\protect\Exp) to the experimental values 
reported by \cite{KrogstadEriksen2013}.}
\label{fig:BT1_Vel}
\end{figure}

\begin{figure}[t]
\centering
\includegraphics[width=0.3\linewidth]{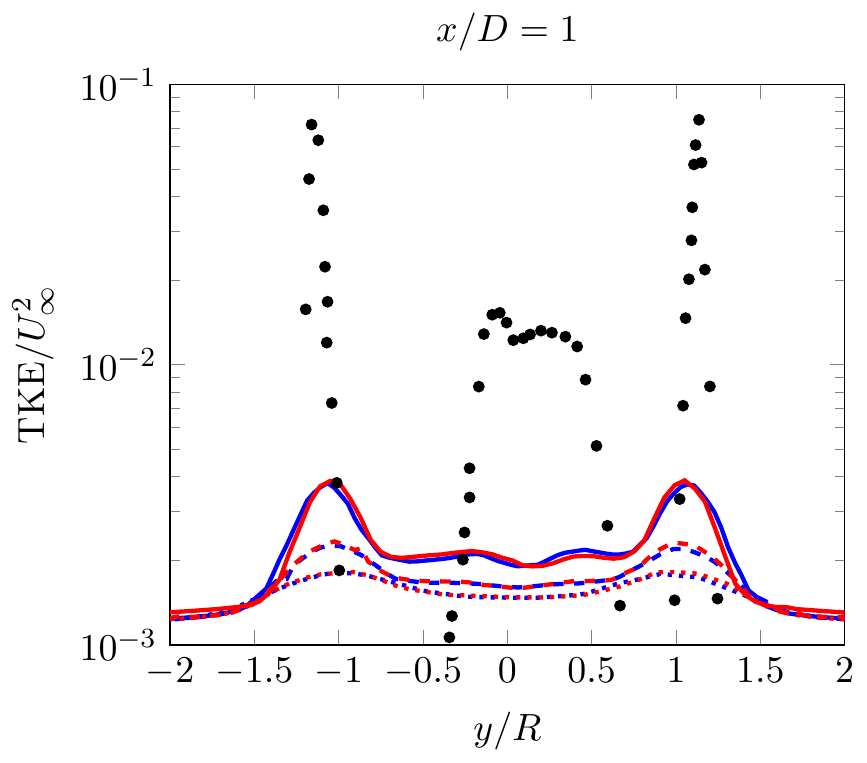} 
~
\includegraphics[width=0.3\linewidth]{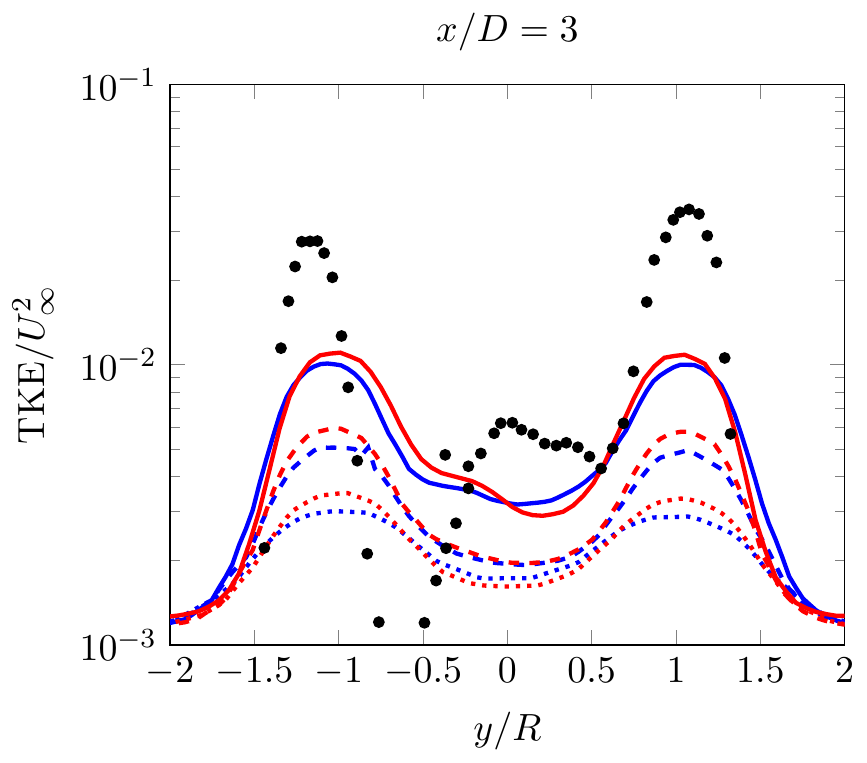} 
~
\includegraphics[width=0.3\linewidth]{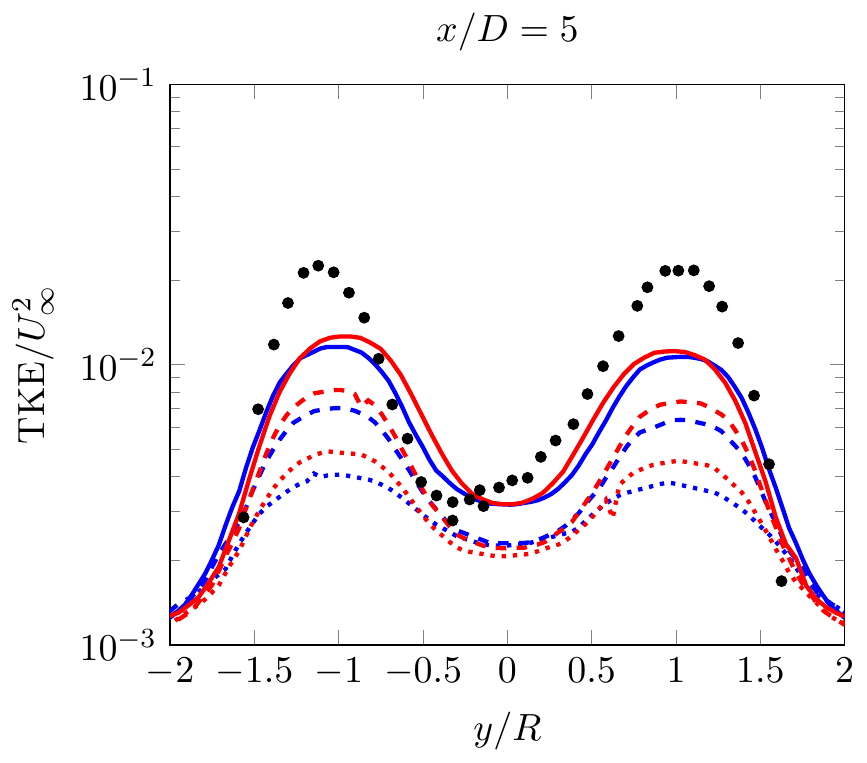} 
\caption{Blind Test 1: Horizontal TKE profiles at $x/D=1,3$ and $5$. Line colouring the same as in figure \ref{fig:BT1_Vel}.} 
\label{fig:BT1_TKE}
\end{figure}

\begin{figure}[t]
\centering
\includegraphics[width=0.3\linewidth]{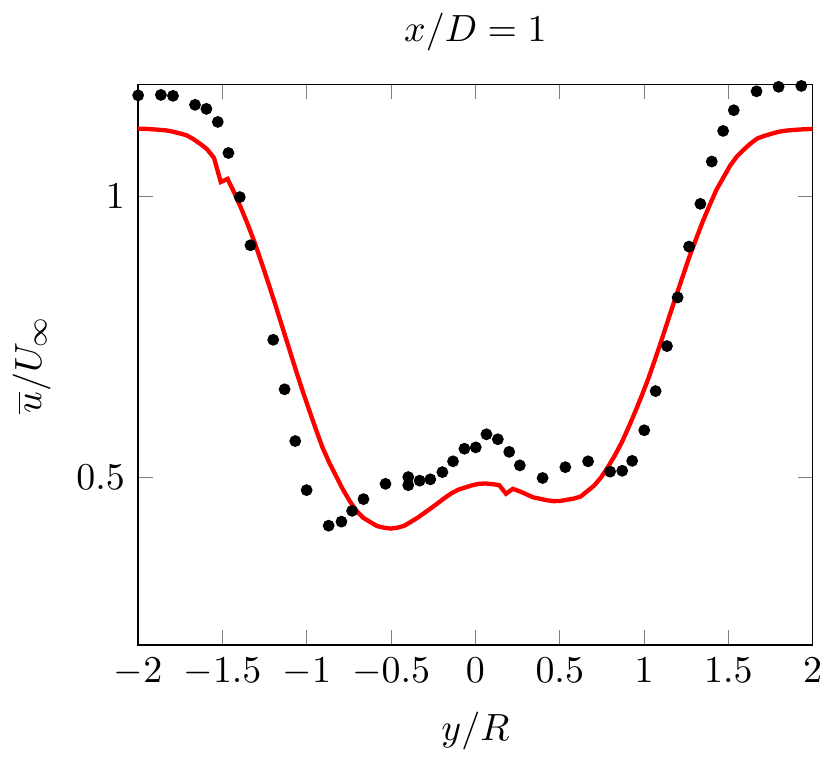} 
~
\includegraphics[width=0.3\linewidth]{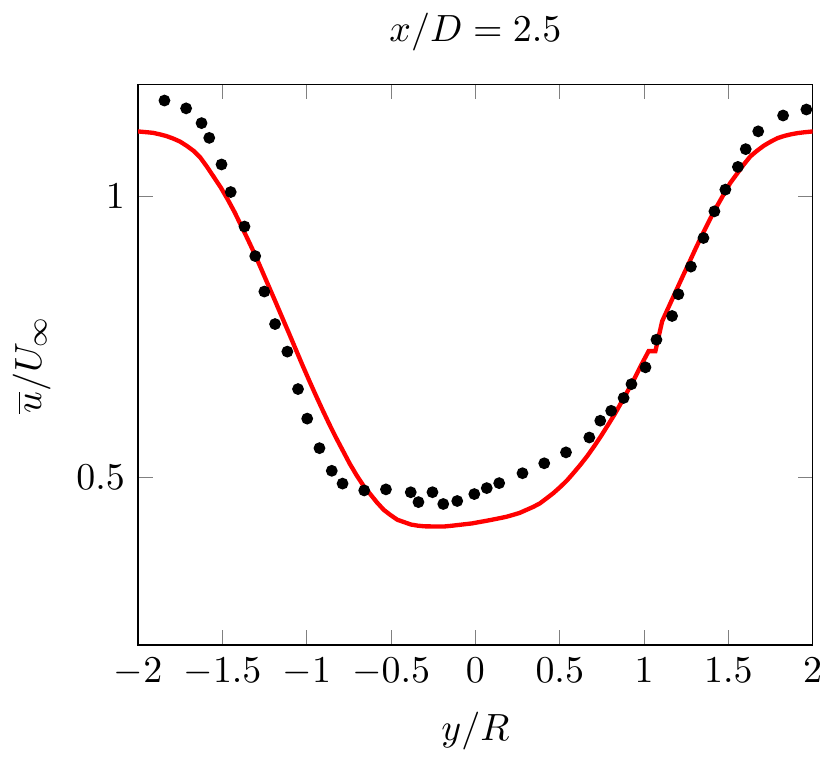} 
~
\includegraphics[width=0.3\linewidth]{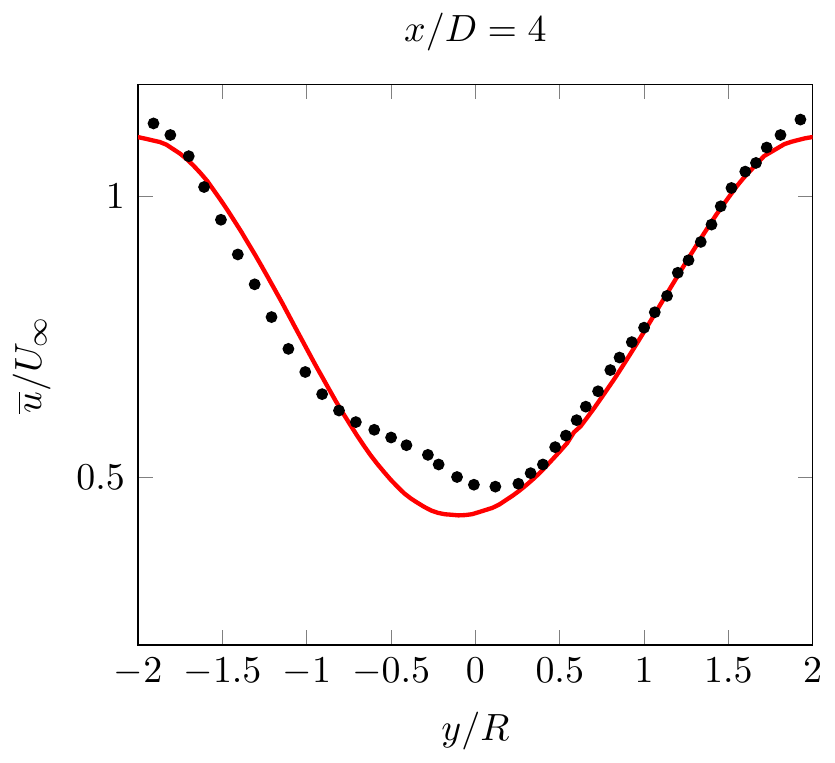} 
\caption{Blind Test 2: Horizontal mean stream-wise velocity profiles at 
$x/D=\,$\numlist{1;2.5;4}. Only the adaptive fine (\protect\AF) solution and the experimental data (\protect \Exp) are shown.}
\label{fig:BT2_Vel}
\end{figure}

\begin{figure}[t]
\centering
\includegraphics[width=0.3\linewidth]{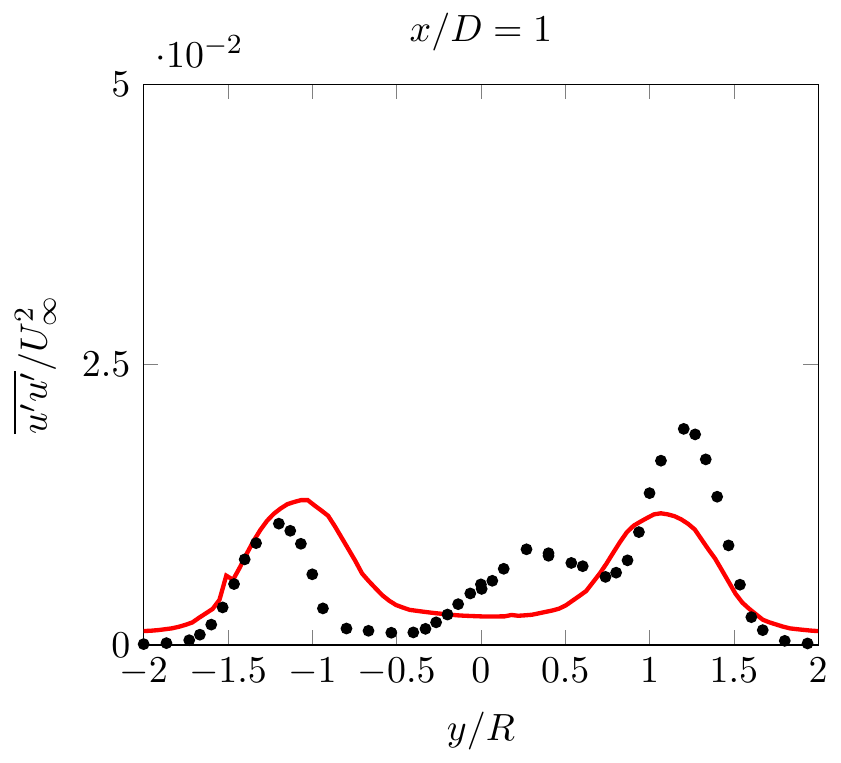} 
~
\includegraphics[width=0.3\linewidth]{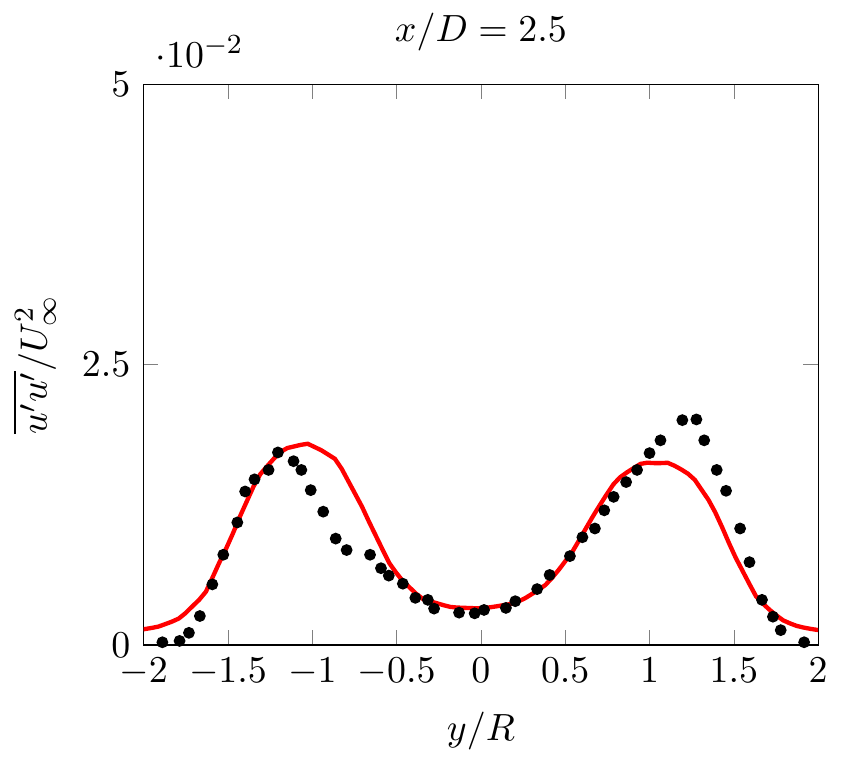} 
~
\includegraphics[width=0.3\linewidth]{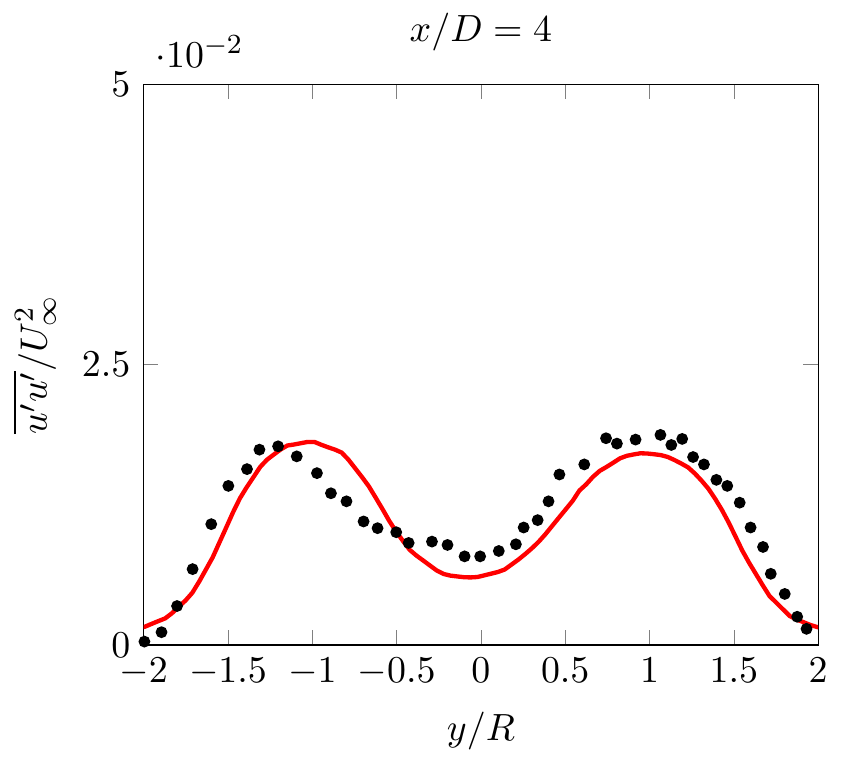} 
\caption{Blind Test 2: Stream-wise turbulent stress profiles at $x/D=\,$\numlist{1;2.5;4}. Only the adaptive fine (\protect\AF) solution and the experimental data (\protect \Exp) are shown.}
\label{fig:BT2_TS}
\end{figure}

\begin{figure}[t]
\centering
\includegraphics[width=0.45\linewidth]{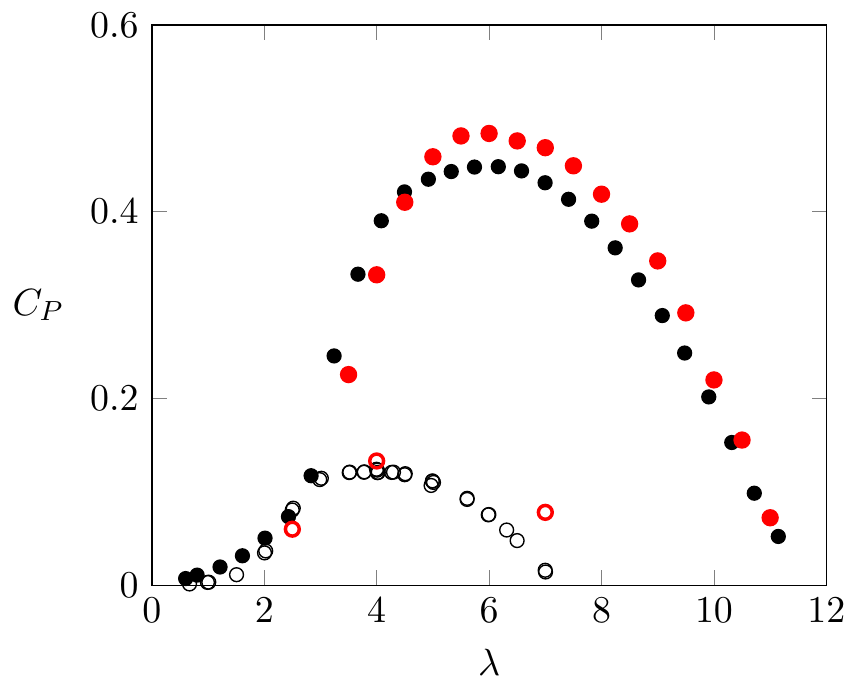} 
\includegraphics[width=0.45\linewidth]{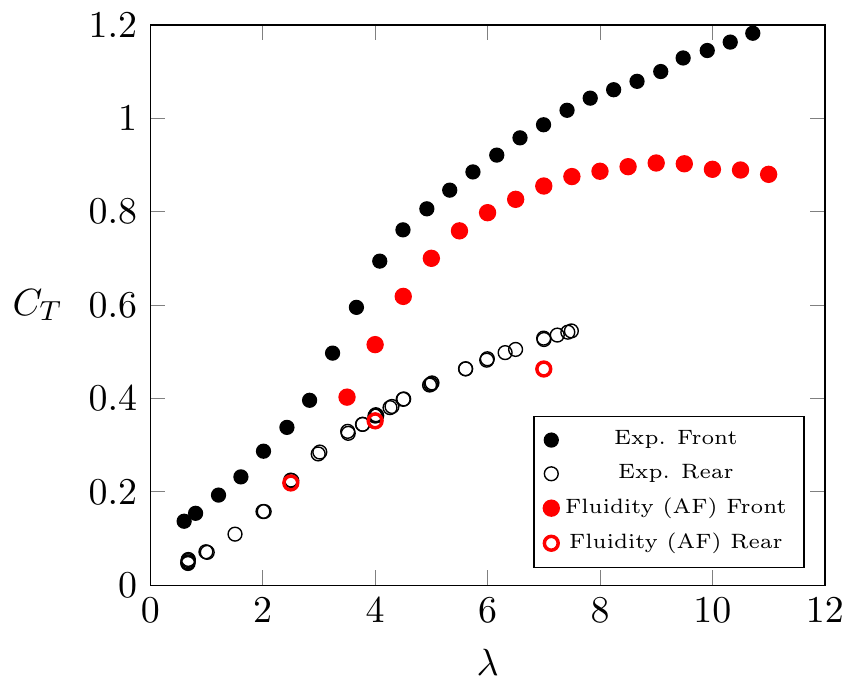}
\caption{Power and thrust coefficient curves as computed by the present model using 
the adaptive fine mesh and as measured by 
\cite{KrogstadEriksen2013,PierellaEtAl2014} plotted against the tip speed ratio $\lambda$.}
\label{fig:BlindTestPerf}
\end{figure}

Having established that the element edge length determines the accuracy of our model we will present the 
wake predictions from BT1 for all six set-up cases described in table \ref{table:cases} using the smallest time step 
$\Delta t=\,$\SI{0.005}{\second}, while results for the power and thrust coefficients as well as the wake predictions of BT2 are presented only the fine mesh-adaptive, in order to maintain clarity.  
Starting with the wake profiles, we present three horizontal profiles downstream of the single 
turbine at $x/D=\,$\numlist{1;3;5} for BT1 and the three horizontal profiles downstream of the 
rear turbine at $x/D=\,$\numlist{1;2.5;4} for only the first scenario ($\lambda_2=\,$\num{4}) 
from BT2. Figures \ref{fig:BT1_Vel} and \ref{fig:BT1_TKE} show the mean stream-wise velocity and 
the TKE for BT1, while figures \ref{fig:BT2_Vel} and \ref{fig:BT2_TS} show the mean stream-wise 
velocity and the stream-wise turbulent stresses $\overline{u'u'}$ for BT2. All quantities are 
time-averaged after a spin-up period which is taken to be approximately \SI{1}{\second}, and have been 
normalised by the upstream velocity $U_{\infty}$ and presented as a function 
of the normalised horizontal distance $y/R$.
Similarly, to resolve the wake field in BT2 only the refined adaptive case was used, and for brevity we present the wake predictions only from 
scenario 1. We should also note that in order to obtain the Reynolds stress $\overline{u'^2}$ we make use of the isotropy turbulence relation $k=3\overline{u'^2}/2$. 
Such an assumption is not appropriate when the turbulence stresses are highly anisotropic, which may explain the large discrepancies in figure \ref{fig:BT2_TS} 
for $x/D=1$. However, better estimates are obtained for the other two downstream profiles. Flow anisotropy was also found to affect the predictions of the wake 
for BT1. This is an inherent inability of all turbulence models, particularly for the estimation of TKE \cite{Pope2000}.   
Looking at integrated rotor quantities such as the power and thrust coefficients, an overall good agreement is observed between the ALM predictions an the wind tunnel measurements. Results
are shown for both BT1 and BT2 in figure \ref{fig:BlindTestPerf}. The only large discrepancy that can be observed in figure \ref{fig:BlindTestPerf} is for scenario 2 of BT2 (rear turbine 
operating with $\lambda=\,$\num{7}). In that case, the power coefficient exhibits a discrepancy of 457.14$\%$ in comparison with the measurements which can be attributed primarily to the ALM 
limitation rather than the modelling of the wake. Indeed, \cite{PierellaEtAl2014} reported this scenario as the most challenging one for turbine parametrisation models. This is due to a 
non-uniform span-wise pressure experienced by the blades of the rear turbine, necessitating the use of blade-resolved simulations to accurately predict the lift and drag coefficients 
of the individual blade elements.
Still, the results from both BT1 and BT2 give us confidence that the proposed adaptive methodology is 
both faster and more accurate than the fixed mesh one. 
However, applying mesh adaptivity in laboratory scale (wind tunnel) experiments cannot 
demonstrate its full potential.
This is due to the high-blockage created in the wind tunnel which necessitates the use of the 
finest mesh resolution almost everywhere in the domain. 
To better demonstrate the potential benefits of mesh adaptivity we present in the next sections simulations for the 
Lillgrund offshore wind farm and discuss some options that can further reduce the computational time.
\section{The Lillgrund offshore wind farm}\label{sec:LillgrundSimulation}
\subsection{Parametrisation of the wind farm}
\begin{figure}[t]
\centering
\includegraphics[width=0.6\linewidth]{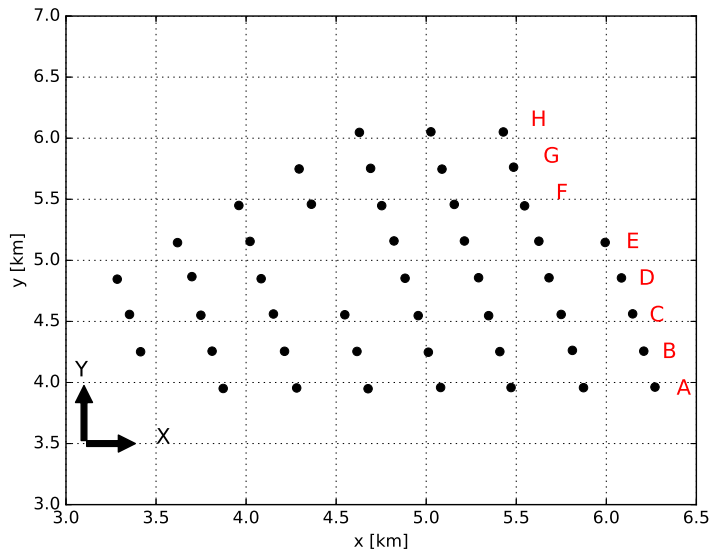}
\caption{Layout of the Lillgrund offshore wind farm. 
 Rows A, B, etc. are aligned with the $x$--axis of the coordinate system.}
\label{fig:Lillgrund_schematic1}
\end{figure}

\begin{figure}[t]
\centering
\includegraphics[width=0.45\linewidth]{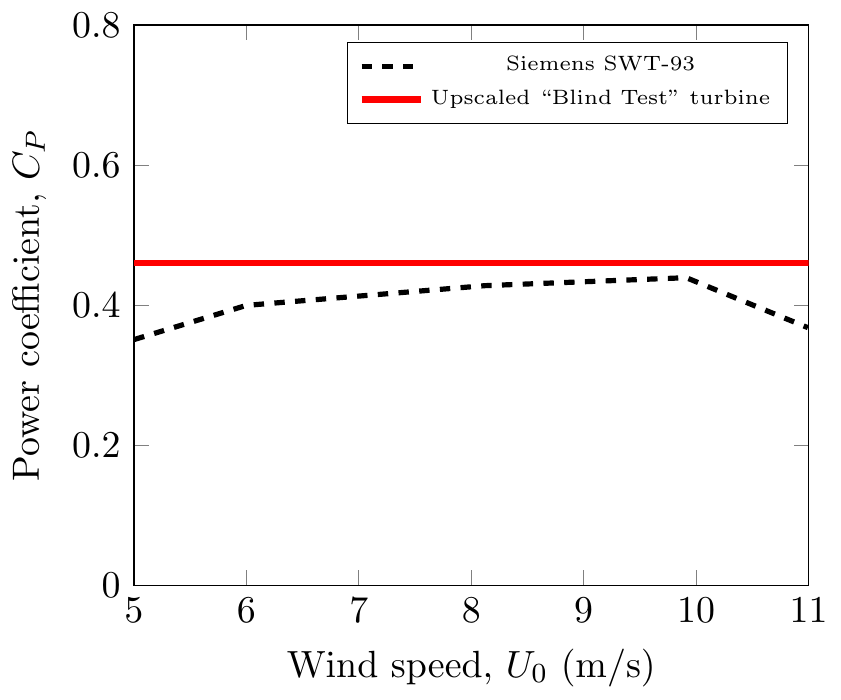}
~
\includegraphics[width=0.45\linewidth]{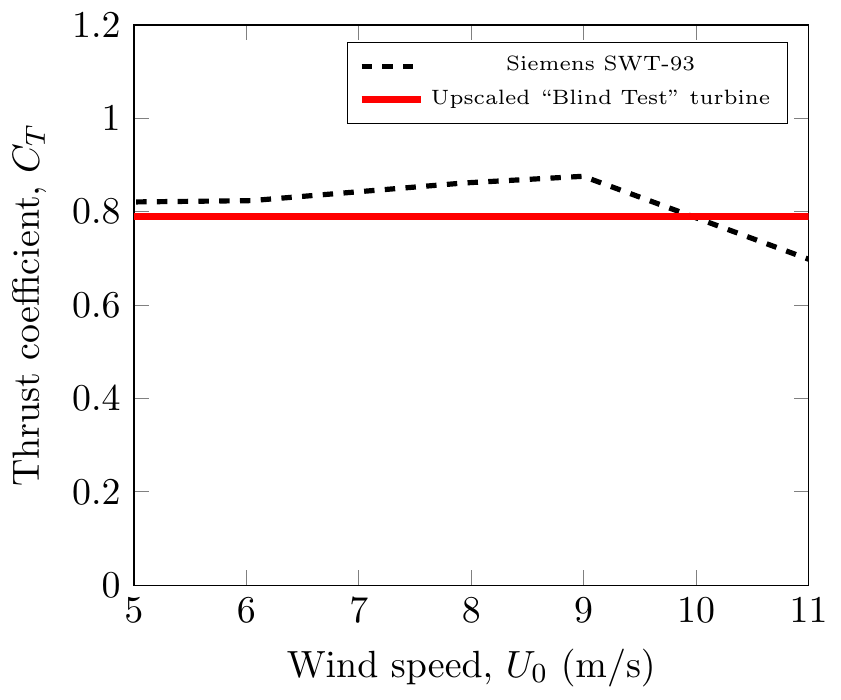}
\caption{Comparison of between the real power and thrust coefficient ($C_P$ and $C_T$) and the up-scaled  ``Blind Test'' turbine as a function of the upstream wind speed.}
\label{fig:UpscaledPerf}
\end{figure}

The Lillgrund offshore wind farm is located near the southern coast of Sweden and has recently attracted the interest of the research community as a benchmark for numeical 
model validation \cite{ChurchfieldEtAl2012a,NilssonEtAl2015,CreechEtAl2015,ErikssonEtAl2015}. Model validation is achieved through comparison
with supervisory control and data acquisition (SCADA) system measurements and it primarily tests the ability of the model to predict power losses along a row of turbines
for different wind directions. Part of the SCADA data are accessible via an online technical report \citep{Dahlberg2009} while the complete information used for the validation 
can be extracted from the validation studies just mentioned, and in particular from the studies of \cite{NilssonEtAl2015,CreechEtAl2015}. The wind farm consists of 48 Siemens SWT93-2.3 MW (Siemens Wind 
Power, Hamburg, Germany) distributed in 8 rows (A to H) as shown in figure \ref{fig:Lillgrund_schematic1}. For our simulations, we consider the Southwestern statistically dominant wind direction 
(\ang{43} / \ang{222}) and we align the coordinate axis system ($x$-$y$) with the rows A, B etc resulting in a layout where the aligned row turbines have a 4.3 rotor diameter spacing.
To parametrise the turbines making up the wind farm we need to make a number of assumptions as their exact blade geometry and airfoil characteristics are not publicly available. 
The available information does include the turbines' rotor radius $R=\,$\SI{46.5}{\meter}, the hub height $H_{\textnormal{hub}}=\,$ \SI{65}{\meter} as well as the thrust and power 
coefficients as a function of wind speed \cite{Dahlberg2009}. For the rest of the turbine parameters we adopt the approach of \cite{NilssonEtAl2015} who considered a downscaled 
version of the conceptual NREL 5 MW turbine as presented by \cite{JonkmanEtAl2009} and confirmed its suitability by comparing the thrust and power (through torque) 
output for different wind speeds. Here, however, we consider an up-scaled version of the ``Blind Test'' turbine, scaled up by approximately a factor of \num{100}. 
To confirm the validity of our choice, we present in figure \ref{fig:UpscaledPerf} the thrust and power coefficients for velocities varying from \SI{5}{\meter/\second} to 
\SI{11}{\meter/\second}, similar with \cite{NilssonEtAl2015}. So far in section \ref{subsec:BlindTests}, we have expressed both the thrust and power coefficient as a function of 
the tip speed ratio $\lambda$. As an active control strategy, we assume that the turbine will simply adjust its angular velocity in accordance with the optimum tip speed ratio, which was found to be around $\lambda=\,$\num{6} as shown in figure \ref{fig:BlindTestPerf}. This assumption is justified in part by the good agreement between the upscaled ``Blind Test'' turbine and the Siemens SWT-93 manufacturer's curve shown in figure \ref{fig:UpscaledPerf}. Here, it is worth emphasizing that a recent investigation by \cite{DeskosEtAl2017} showed that when wake predictions are made the need for mesh resolution is primarily driven by the velocity deficit, which inherently relates to the thrust coefficient. Therefore any turbine 
parametrisation that accurately captures the thrust force would result in a similar computational mesh. 
\subsection{Mesh strategies}
For the simulations presented here two mesh strategies are investigated, namely the fixed (but pre-refined, i.e. variable resolution) and fully dynamically adaptive meshing. 
Considering that the problem at hand is inherently multi-scale and a number of length scales, spanning from the individual turbine wake's turbulence length scale to the far 
larger atmospheric meso-scales usually in the order of $\mathcal{O}($\SI{200}{\meter}$)$, need to be resolved, the two selected meshing strategies are the following:
\begin{enumerate}
\item a large domain of dimensions \SI{10}{\kilo\meter} $\times$ \SI{10}{\kilo\meter} $\times$ 
\SI{1}{\kilo\meter} is considered with a meso-scale resolution $\mathcal{O}($\SI{200}{\meter}$)$ everywhere expect for an inner region with prescribed resolution using an edge length equal to 
$h=\,$\SI{20}{\meter} (in the subsequent discussions this is referred to as the pre-refined region -- PRR) and, \item a large domain of the above mentioned dimensions and meso-scale initial 
resolution everywhere without the a priori chosen inner region refinement, but with mesh-adaptivity enabled and operating over the whole domain and a minimum allowed edge length of 
$h=\,$\SI{20}{\meter} specified.
\end{enumerate}
\begin{figure}[t]
\centering
\includegraphics[width=0.85\linewidth]{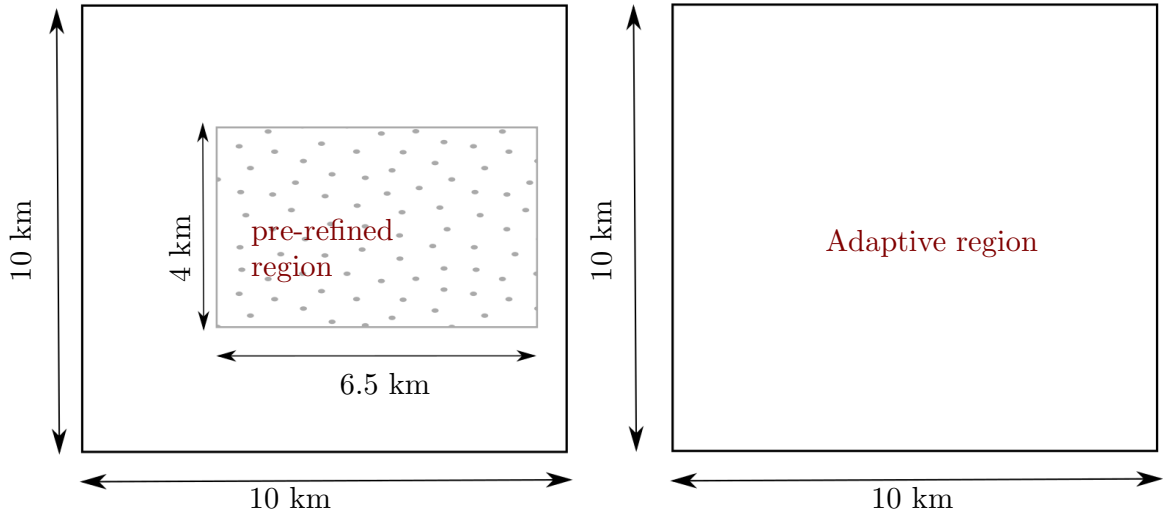}
\caption{Schematic representation of the two mesh strategies considered for the Lillgrund offshore wind farm test case. The pre-refined region
is centred around the middle of the $y$-axis while it starts \SI{3}{\kilo \meter} from the domain inlet and ends \SI{0.5}{\kilo \meter} before
its outlet.}
\label{fig:Mesh_Strategy}
\end{figure}
The simulations were performed for a case with a pre-refined region (PRR) assigned in an area of \SI{4}{\kilo\meter} $\times$ \SI{6.5}{\kilo\meter} $\times$ \SI{0.16}{\kilo\meter} 
(see fig. \ref{fig:Mesh_Strategy}), and three adaptive simulations in which different frequencies of mesh adaptations ($T_{\textnormal{adapt}}=\,$\SIlist{2.5;5;10}{\second}) were applied. The 
selection of the pre-refined region (PRR) was performed through a ``trial and error'' approach as the width and the height of the individual turbine wakes were not known a priori. Therefore 
our goal was to create a domain which has the refined region covering the wake region but does not extend excessively beyond this point. An alternative approach (which would arguably represent 
a fairer comparison to the adaptive approach) would be to assume no a priori knowledge and simply use the minimum edge length over the entire domain, but this would 
of course result in a huge problem size. For the mesh adaptive simulations the adaptation period is selected from a larger number of user-defined parameters available to us such as the 
elements' maximum aspect ratio, the minimum/maximum element ratio etc. This option is considered in order to understand its impact on reducing the overall CPU time in the present uRANS set-up. 
The selection of mesh adaptivity frequency as a key parameter is motivated by the computational cost (runtime penalty) associated with it, particularly during the interpolation process. 
Indeed, mesh adaptations can be seen as a potential bottleneck step in the computations, as load-rebalancing represents a data migration overhead and the application of relatively 
costly interpolation methods to be used (e.g. Galerkin projection) to transfer the information from the previous mesh to the new one. The application of a Galerkin projection method is 
necessitated here by the use of a discontinuous function space, as the recent study of \cite{FarrellMaddison2011} showed that a consistent interpolation is not suitable for discontinuous 
fields. For the present simulations, we consider the wind speed to be constant and therefore the required frequency of the mesh adaptation is not expected to vary with time, except for when 
the wakes have converged. However, if longer simulations are to be undertaken in which the wind speed is expected to vary with time, the frequency of the adaptations may be adjusted during the 
course of the simulation to better capture the wake dynamics. It should be noted of course that a varying wind direction would also necessitate a larger PRR in the fixed mesh case, leading to 
substantial additional computational costs. Finally, other parameters such as the element edge length size $h$, have been excluded as in section \ref{sec:ValVer} we showed that there is a 
direct correlation between the size of the underlying mesh and the model's accuracy. Thus, changing the frequency with which these adaptations occur is a key remaining parameter which can be 
varied in order to reduce the overall CPU time while maintaining a similar accuracy for the turbine performance and wake predictions. The latter will also be re-assessed later in this section.
\subsection{Simulation setup} 
\begin{figure}[t]
\centering
\includegraphics[width=\linewidth]{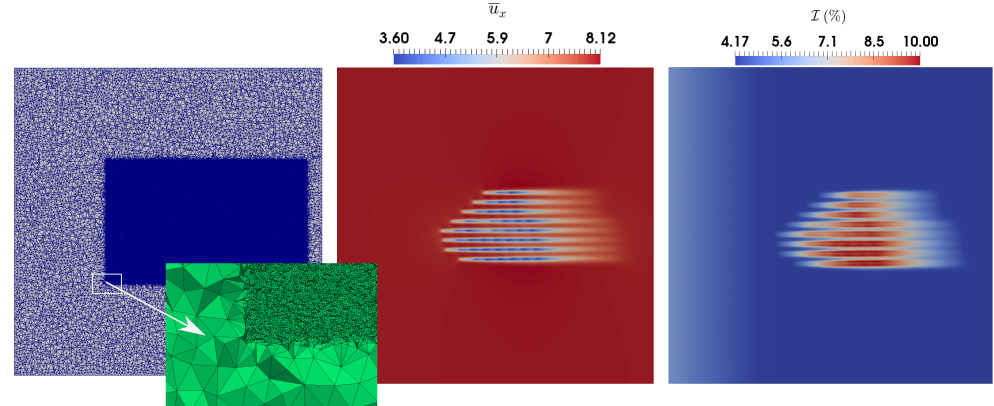}
\caption{Fixed mesh simulations: From left to right a horizontal cross-section at hub height of the underlying mesh, the stream-wise velocity $u_x$ 
and the turbulent intensity $TI$ are shown.}
\label{fig:FixedMesh_Solution}
\end{figure}

\begin{figure}[ht]
\centering
\includegraphics[width=\linewidth]{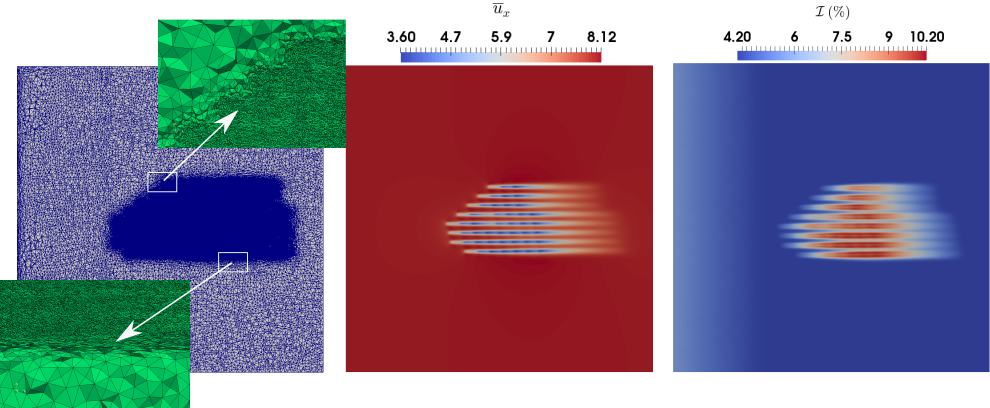}
\caption{Adaptive mesh simulations ($T_{\textnormal{adapt}}=\,$\SI{2.5}{\second}): From left to right a horizontal cross-section 
at hub height of the underlying mesh, the stream-wise velocity $u_x$ and the turbulent intensity $TI$ are shown.}
\label{fig:AdaptiveMesh_Solution}
\end{figure}

\begin{figure}[t]
\centering
\includegraphics[width=0.4\linewidth]{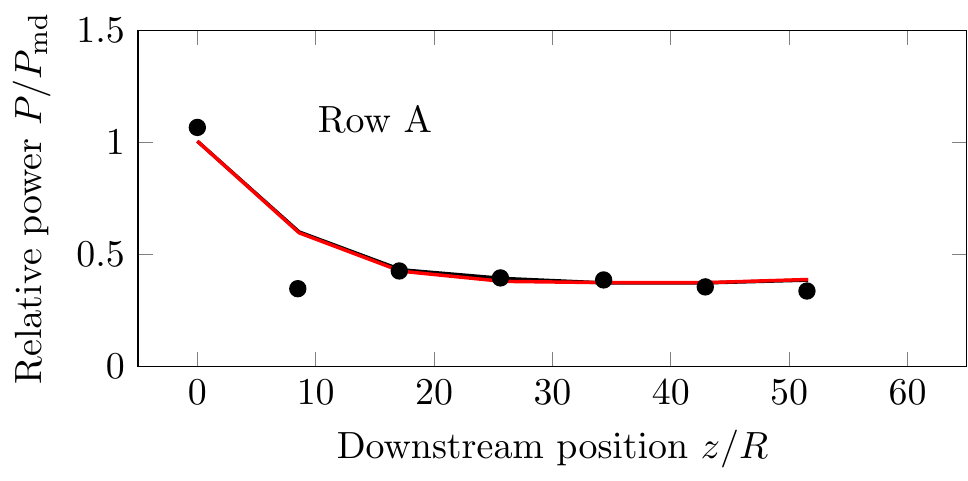}
~
\includegraphics[width=0.4\linewidth]{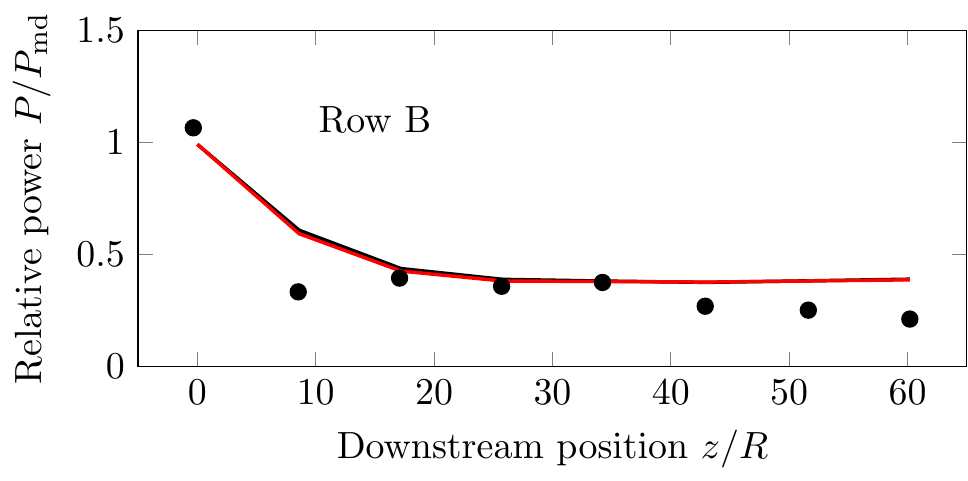}
\\
\includegraphics[width=0.4\linewidth]{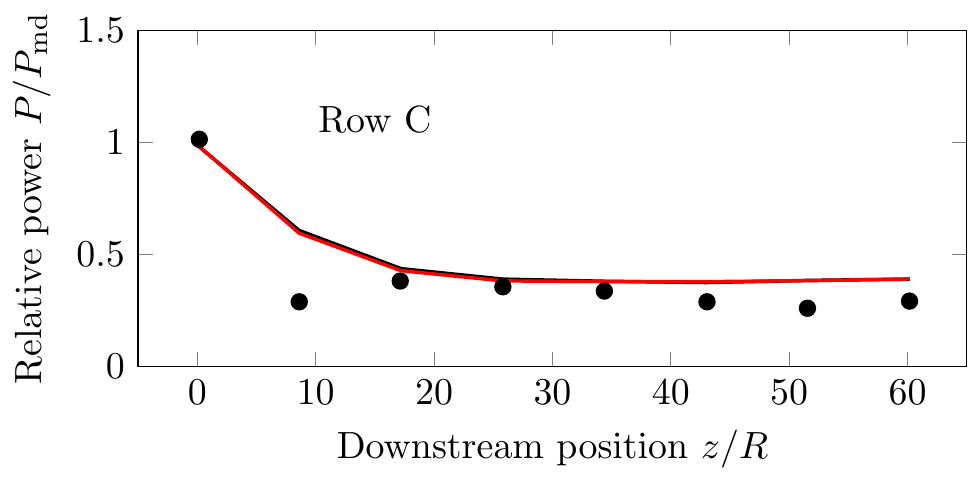}
~
\includegraphics[width=0.4\linewidth]{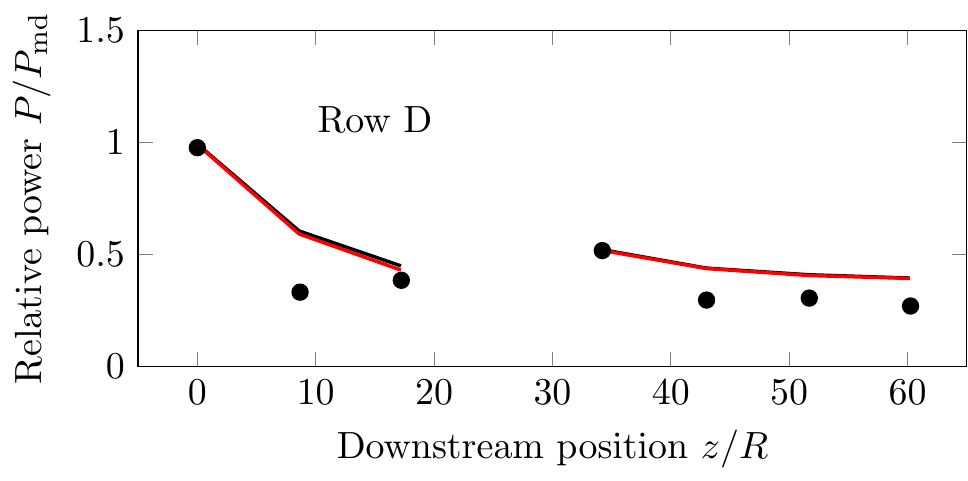}
\\
\includegraphics[width=0.4\linewidth]{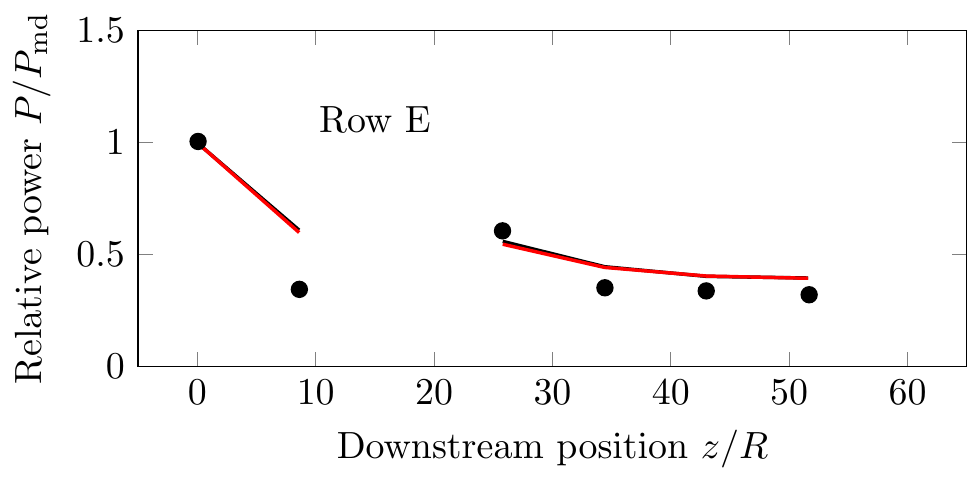}
~
\includegraphics[width=0.4\linewidth]{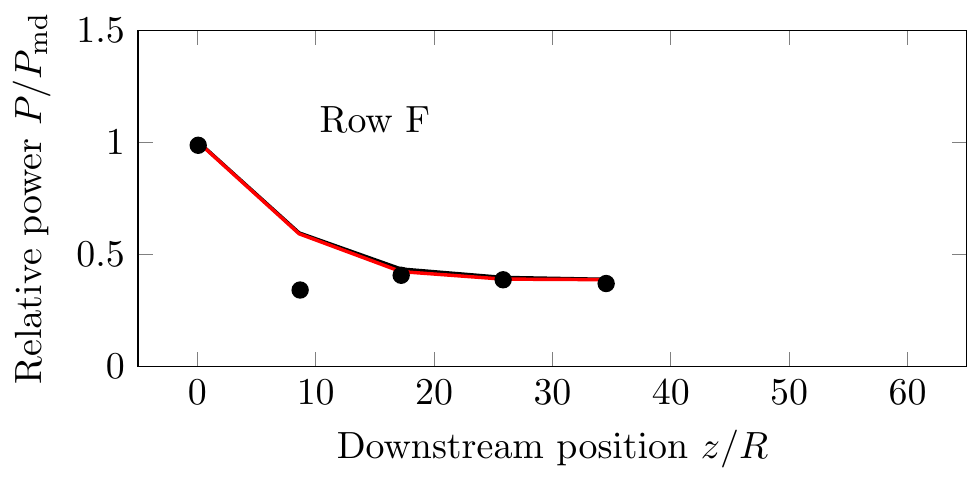}
\\
\includegraphics[width=0.4\linewidth]{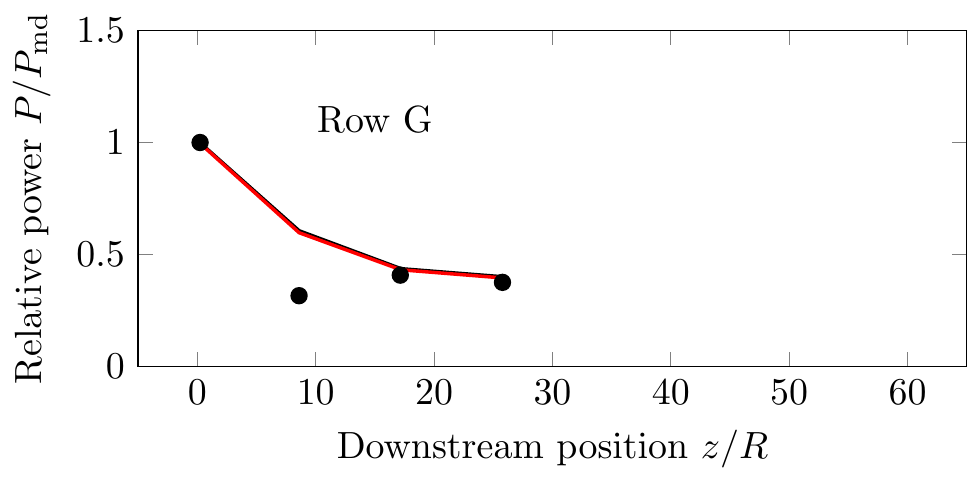}
~
\includegraphics[width=0.4\linewidth]{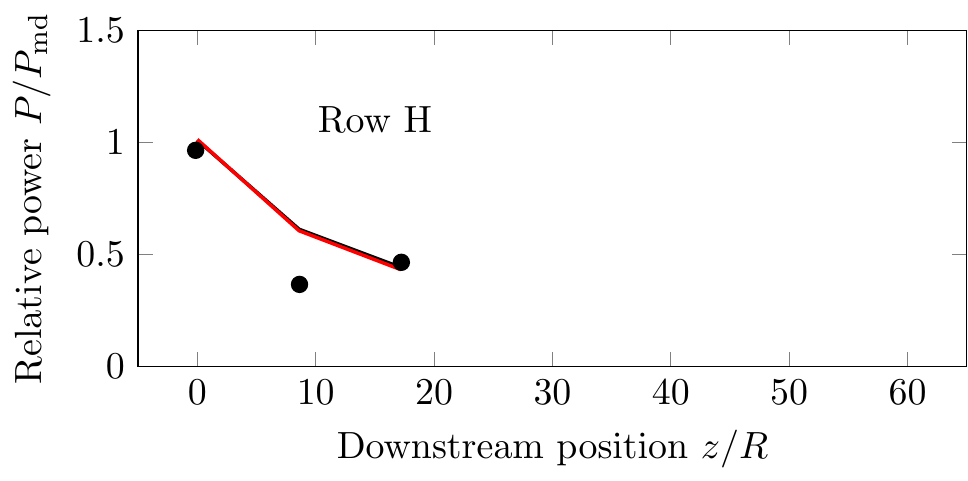}
\caption{Normalised power computed by the fixed-mesh (black line) and the mesh-adaptive (red line) simulations for turbines in Row A--H and 
compared against the measured data (dots) of \cite{Dahlberg2009}}
\label{fig:Relative_Power_Estimates}
\end{figure}

Moving on to the simulation set-up, for the purpose of these simulations we assume uniform initial and inlet mean velocity and TKE profiles ($U_0=\,$\SI{8}{\metre \per \second} and 
$k=\,$\SI{0.31}{\metre^2 \per \second^2} which corresponds to a turbulence intensity $\mathcal{I}=\,$\num{5.7}$\%$). On all other boundaries, we apply free-slip velocity and zero gradients 
for all other quantities ($\partial/\partial n(u,v,w,k,\omega)=\,$\num{0}). The selected inlet and boundary conditions considered herein are not representative of the levels of shear that the 
turbine wakes will experience within a realistic atmospheric boundary layer. However, as a first approach, we will demonstrate the benefits of mesh adaptivity by imposing a uniform incident 
velocity and TKE profile, and therefore based on the selected adaptation metrics (velocity and TKE) not need to refine outside the wake region. The accuracy of our choice is also discussed 
later, when the numerical results are compared with the observed SCADA data and some discrepancies are observed. It should be remembered that the focus of this study is to compare the PRR solutions with the mesh-adaptive ones, and the observed data is added only to show the overall performance of the model. Future studies will consider more complex scenarios in which the mesh will also be further controlled in a manner that allows for anisotropic gradation in the vertical direction only, outside the wake region. Nevertheless, to obtain a 
quasi-steady solution for the far wake field we time-march the solution with a constant time step of $\Delta t=\,$\SI{0.1}{\second} and the solution 
for a total number of \num{4000} time steps. The final wake solutions are shown in figures \ref{fig:FixedMesh_Solution} and 
\ref{fig:AdaptiveMesh_Solution}, for both the fixed and adaptive mesh (using the smallest adaptation period $T_{\textnormal{adapt}}=\,$\SI{2.5}
{\second}) simulations. In addition, the power estimates normalised by the median of the production of the front row turbines $P_{\textnormal{md}}$ (similar to quantities considered by 
\cite{NilssonEtAl2015}) are presented and compared against the measured SCADA data in figure \ref{fig:Relative_Power_Estimates}.
The two approaches' (adaptive and fixed mesh) results exhibit very similar behaviour both qualitatively and quantitatively  as observed in figures \ref{fig:FixedMesh_Solution} and \ref{fig:AdaptiveMesh_Solution}. An important feature in both solutions is the dissipation of the inlet TKE due to the assignment of a turbulence frequency required for the dissipation of the 
turbine wakes. The increased dissipation creates a moving front for the inlet TKE which however disappears after about \num{2000} time steps. This impacted on the adaptive mesh since the TKE 
field was used within the definition of the metric controlling mesh optimization. The impact of the TKE inlet front is discussed in more detail in the next subsection. Next, the power 
production trends are also found to agree well with measurements as shown in figure \ref{fig:Relative_Power_Estimates}. The only major discrepancy observed in the plots (Rows A--H) 
is the over-prediction of the relative power of the second turbine. This discrepancy may be attributed to our selection of uniform inflow conditions, as the amount of computed shear and 
wake asymmetry will be underestimated. Similar trends were also observed by \cite{ChurchfieldEtAl2012a,NilssonEtAl2015,CreechEtAl2015} who adopted a log--law profile, 
although the discrepancies between the experimental values and their LES results appear to be much smaller.  
Nevertheless, the power production as predicted by the three adaptive simulations at the final time level are essentially identical and therefore only 
one of the three is plotted in figure \ref{fig:Relative_Power_Estimates}. The mean relative error over all turbines between the CFD simulations' power prediction and the 
measured data amounts to \num{7.38}$\%$ for all three adaptive-mesh simulations while it takes the value \num{8.02}$\%$ for the fixed-mesh ones. This result confirms our 
initial hypothesis that varying the frequency of mesh adaptations will not affect the accuracy of the model predictions in the case of the uRANS approach employed here. On the other hand, a 
small difference between the fixed-mesh and the adaptive-mesh simulations' power predictions is observed; this can be attributed to the ability of the adaptive simulations to better resolve
the wake field by applying an optimum element aspect ratio, as was also shown for the wind tunnel tests. 
\subsection{Computational efficiency}\label{subsec:CompEff}
\begin{figure}[t]
\centering
\includegraphics[width=0.45\linewidth]{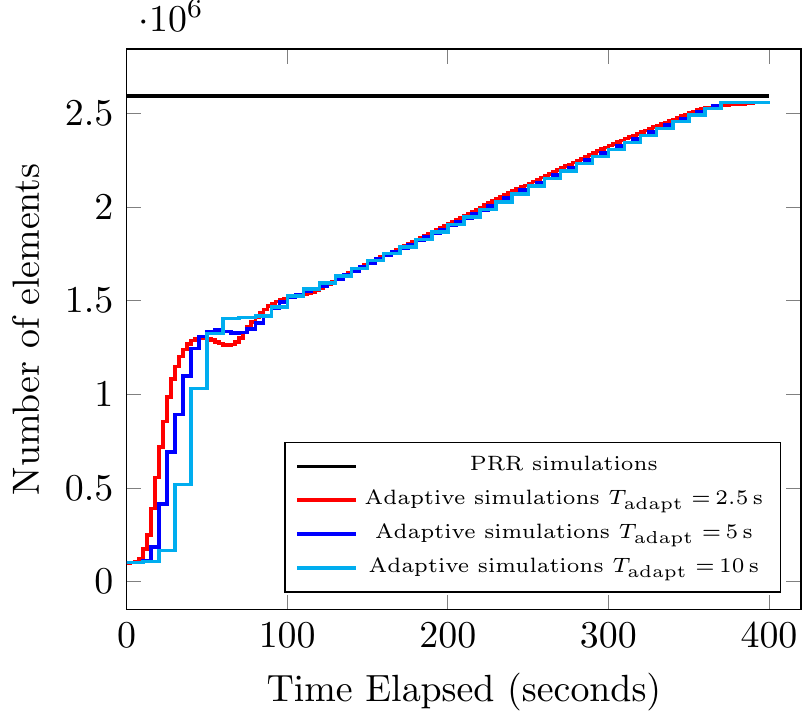}
~
\includegraphics[width=0.45\linewidth]{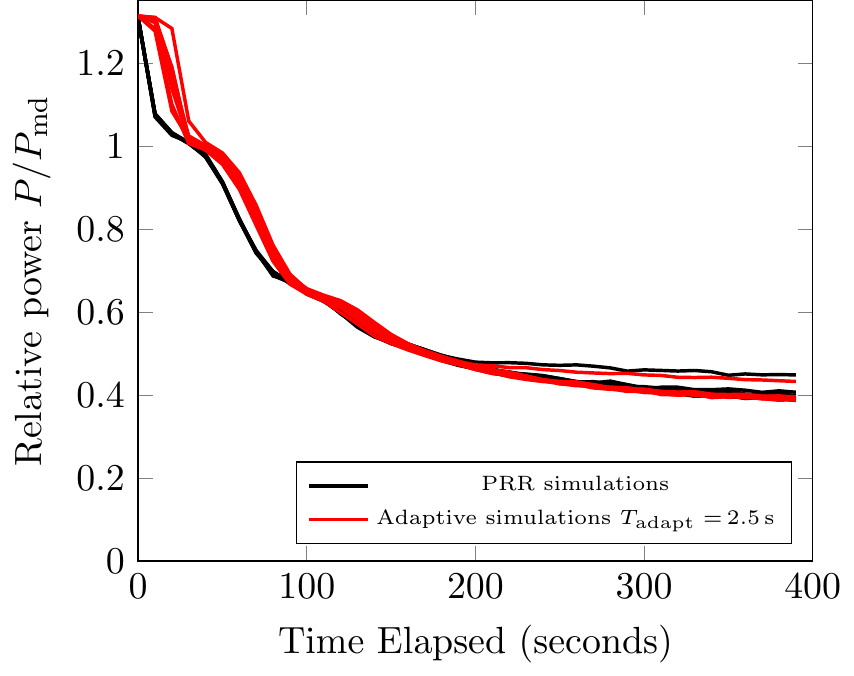}
\caption{Left: Number of tetrahedral elements used for the underlying mesh as a function of the 
elapsed simulation time, Right: Convergence of the relative power of the rear row of turbines -- individual lines for each turbine are plotted here.}
\label{fig:ElementsCount}
\end{figure}

\begin{figure}[t]
\centering
\includegraphics[width=\linewidth]{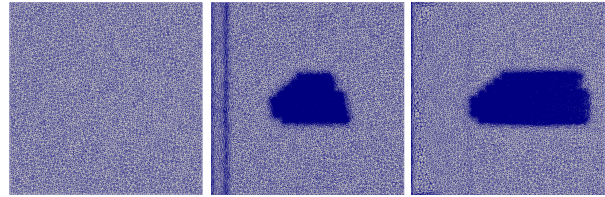}
\caption{From left to right: A horizontal slice at hub height ($H=\,$\SI{65}{\meter}) through the adaptive mesh at \SIlist{0;100;400}{\second}, and having in total \numlist{100760;1649459;2558880} elements in the computational domain, respectively.}
    \label{fig:AdaptiveMesh_HorizontalSlice}
    \end{figure}

\begin{figure}[t]
    \centering
    \includegraphics[width=0.9\linewidth]{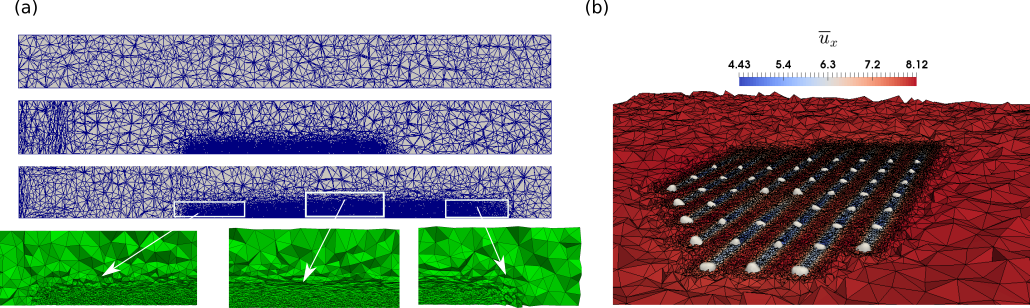}
    \caption{Left: A vertical slice along the $x$-axis passing through the centre of the domain ($y=\,$\SI{5}{\kilo\meter}) for the adaptive mesh at \SIlist{0;100;400}{\second} (from top to bottom) having in total \numlist{100760;1649459;2558880} elements in the computational domain, respectively, and zoomed--in snapshots from the final adaptive mesh ($t=\,$\SI{400}{\second}) at
    different locations. Right: A 3D view of the computational mesh and stream wise velocity $\overline{u}_x$ at $t=\,$\SI{100}{\second} using a crinkle slice passing through hub height. Iso--contours of the magnitude of the turbine source term $F_T$ are also shown to identify the location of each turbine. The iso--contour level is chosen small enough such that all turbines are visible.}
    \label{fig:AdaptiveMesh_VerticalSolution}
    \end{figure}

Returning to a key objective of our investigation, which is to examine the computational efficiency of the two proposed mesh strategies, it was hypothesized that the computational cost of
the adaptive simulations should be lower than that for the fixed pre-refined mesh simulations, as well as the computational cost becoming smaller as the adaptation period is increased. In 
addition, the same or similar accuracy should be achieved when the same minimum edge length is used. To test our hypothesis, we conducted simulations using one fixed and three adaptive meshes. 
The simulations were run in parallel using MPI on a total of 80 processing cores (4 nodes each of 20 cores) on the cx1 cluster at Imperial College London. The pre-refined simulations required 
approximately 86 hours to complete which is considered as a reference value. We should mention here that the absolute value of the required CPU time would differ for different 
algorithms or code implementations. For this reason, we present the CPU times for the adaptive mesh simulations normalised by the CPU time for the pre-refined mesh case. Here the term ``CPU time'' refers to the wall--clock time multiplied by the number of processing cores. This CPU time is dominated by the actual run time of the jobs and auxiliary computational 
procedures such as the decomposition and re--partition of the mesh were found to result in trivial computational times in comparison. With this in mind, we start by presenting the temporal evolution of the number of fluid mesh elements in figure \ref{fig:ElementsCount}; the number of elements are seen to follow a similar trend for all three adaptive cases.
The simulations begin with a spin-up of the mesh in the vicinity of the turbines and a moving front starting from the inlet. The latter, is a result of the dissipation of the turbulent kinetic 
energy in the inlet and disappears after about \SI{50}{\second}. After this spin-up period, and between \SIrange{50}{100}{\second}, the number of elements are slightly reduced due to the 
``dissipation'' of the moving front, after which the element counts start to increase again with a constant rate after \SI{150}{\second}. Finally, the number of elements becomes constant after 
approximately \SI{380}{\second} at which time a steady state solution has been reached for the far field. A similar trend for the count of elements/DoF was also obtained by 
\cite{KirbyEtAl2017}. In their simulations of the full Lillgrund wind farm they observed a strong linear spin--up curve due to wake transients, and subsequently a flattened peak where the 
wakes have started interacting with each other, and small variations in the mesh due to wake--wake and turbine--wake interactions were found. Such fluctuations are not observed however in the 
dynamic mesh evolution of the present simulations, which can be attributed to the low--pass temporal filtering incurred by the uRANS equations. On the right hand side of the same figure, we 
have plotted the evolution of the relative power ($P/P_{\textnormal{md}}$) for all the back-row turbines against the elapsed time, for one mesh-adaptive case ($T_\textnormal{adapt}\,=$\SI{2.5}
{\second}) and the PRR simulation. It can be observed that the relative power of the individual turbines converges to near its final value long before \SI{380}{\second}. The evolution of the 
adaptive mesh is presented via horizontal and vertical slices through the domain in figures \ref{fig:AdaptiveMesh_HorizontalSlice} and \ref{fig:AdaptiveMesh_VerticalSolution}. 
Note that slices through completely unstructured tetrahedral meshes do result in complex polygons and slivers, but these images do convey the spin up of the mesh 
as it resolves the developing wakes. Zoomed pictures of the vertical profiles are also shown in \ref{fig:AdaptiveMesh_VerticalSolution} to demonstrate the element gradation (particularly the higher aspect ratio of the elements) near the edge of the wake field.
Element--wise, the plots (figure \ref{fig:ElementsCount}) of the three element counts for the adaptive mesh simulations remain well below that for the pre-refined simulation line until the end 
of the simulations and thus we may argue that since the adaptive-mesh simulations always use a smaller number of elements than the fixed-mesh ones, and that the Fluidity solver 
scales approximately linearly with the number of elements, then the adaptive-mesh simulations should require a much lower CPU time. For the adaptive mesh simulations, however, additional CPU 
``penalties'' are imposed due to the various stages of the mesh adaptation procedure itself. Thus, the resulting CPU time can be significantly impacted by the frequency of these mesh 
adaptations. In addition, there is also potential for unequal balancing of the elements across the MPI processes (although a dynamic load balancing step is 
incorporated within the parallel mesh optimization procedure). The fact that a fixed number of processing cores is utilised throughout the simulation, and that this number is unlikely to be 
optimal from a parallel scaling perspective given the significant changes in element count during the course of the adaptive mesh simulations, can also impact on computational efficiency 
gains. The actual CPU required by our numerical simulations confirm our initial hypothesis. For the pre-refined simulations a total of 
\num{6880} CPU hours were required, while \numlist{5641.6;4974.24;4747.19} CPU hours were required for the adaptive mesh simulations using mesh adaptation periods of 
($T_{\textnormal{adapt}}=\,$\SIlist{2.5;5;10}{\second}), respectively. From these results, we may make two observations. First, by switching to the adaptive 
approach the CPU time is reduced by \num{18}$\%$ even when making relatively frequent mesh adaptation operations. Second, by reducing the 
frequency of mesh adaptations additional reductions in the CPU time are achieved: to \numlist{27.7;31} $\%$ respectively. However, we should 
mention that the selection of \num{80} processors while it can be optimal in terms of scalability in the case of the pre-refined region,
it is not always so for the adaptive ones. This is due to the fact that during the spin-up period, a relatively small number of elements will
be distributed across a large number of processors and the computational cost will be dominated by the MPI communications between the 
partitions. 
\begin{figure}[t]
\centering
\includegraphics[width=0.45\linewidth]{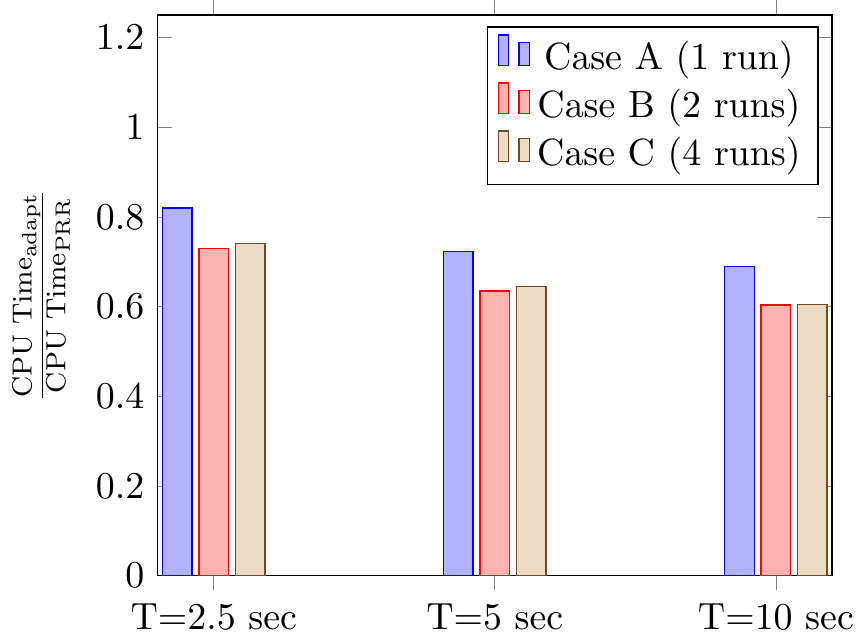}   
\caption{CPU time required by the adaptive simulations in the three scenarios (A,B and C), normalised by the CPU time required for the PRR simulation.}
\label{fig:CPU}
\end{figure}
In order to further optimize the CPU usage of the adaptive simulations, we have re-run the same simulations but this time starting the parallel computation on a smaller number 
of processing cores (Nproc=\num{20}), and as the overall problem size grows we start and stop the simulation (using checkpoints) and increase the number of cores to 
the final \num{80} as the element counts increase. We should emphasize here that prior knowledge of the final mesh from the original adaptive simulations using \num{80} processing cores was 
essential in better designing the decomposition of the domain when checkpoints were used. Here we present three cases, the first one (Case A) is the original case in which, the simulations 
were run on 80 processors for the whole time period, the second one (Case B) starts with \num{20} processors, then after the first spin--up time ($t=\,$\SI{100}{\second}) the simulation is 
stopped and restarted with \num{80}, while the last one (Case C) uses \numlist{20;40;60;80} at the periods \numrange{0}{100}, \numrange{100}{200}, \numrange{200}{300} and \numrange{300}{400}\SI{}{\second}, respectively. The cumulative CPU time for the three cases is shown in figure \ref{fig:CPU}. It is observed that cases B and C result in a smaller overall
CPU time than case A. In addition, while in case C the decomposition/repartition is performed four times, this does not lead to smaller CPU times. 
Instead, case B seems to behave better, although by a small amount. Based on this, we may infer that if a moderately large number of elements ($\simeq\,$\num{25000}) is used per processor 
after the first spin-up time ($t=\,$\SI{100}{\second}), changing the number of processors more often does not lead to significant changes in the observed CPU times. This might be due to the 
fact that Fluidity exhibits excellent scalability properties (both strong and weak) when a small number of processors is used, and enough elements are distributed to each processor.
\section{Discussion}\label{sec:Discussion}

In this work, we presented the implementation and validation of a uRANS-based, mesh-adaptive ALM which is able to optimize the number of elements/cells that are required to resolve the wake 
field within a full-size wind farm. Our ALM implementation is based on the original model of \cite{SorensenShen2002} with additional models being used to 
account for blade end effects and the impact of the tower shadow on the near wake field. The fluid flow is resolved using a uRANS formulation of the governing equations and the 
$k$--$\omega$ SST turbulence closure model. The developed Fluidity-ALM model was validated against wind tunnel measurements before attempting full-scale wind farm simulations. For the wind 
farm simulations we selected the Lillgrund offshore wind farm for which numerous previous studies and observed data are available. The effect of mesh-adaptivity on reducing 
the computational cost was also examined with particularly emphasis being given on processors' load balancing, especially during the transient period where the number of elements follows a 
linear trend.

More specifically, for the developed model a detailed mesh convergence study was undertaken for both the fixed and the adaptive mesh approaches showing that for a given edge length $h$ the two 
mesh strategies yield similar results for the integrated rotor characteristics such as the power or thrust coefficient. Moreover, as the element's edge length is reduced, the model predictions 
for the power coefficient $C_P$ converge to the experimental reference values. Looking at the wake solutions on the other hand, mesh-adaptivity provides better estimates for the wake field, 
particularly in regions where large shear (velocity gradients) exist. This is due to the ability of the mesh optimization algorithm to identify these regions and allow elements with 
large aspect ratios to align with the underlying gradients in solution fields. A fixed mesh does not have such an ability and therefore a better solution at these regions can 
be obtained only by further refining the mesh locally. The validation of the model using the two Blind Tests (BT1 $\&$ BT2) shows that both the near and far wake field can be accurately 
predicted given that enough resolution is used. More satisfactory, however, are the predictions from the second validation case, BT2. Deep array wake modelling has been a great challenge for 
many wake models and it is a crucial step towards accurately predicting offshore wind farm power output. The results from the comparison with BT2 give us confidence that, although only two 
turbines in-line are used, we are capable of obtaining high-fidelity wake solutions for the back row turbines, when both operate at peak conditions (optimal tip speed ratio).

Looking at the Lillgrund offshore wind farm simulation results, the power predictions from each row (A--H) agree well with measured data from  \cite{Dahlberg2009}.
The largest discrepancy between the model predictions and the measured data is observed on the second turbine in each row. This systematically appears in all simulations, and is believed 
to be due to the torque chosen for the second turbine. Nevertheless, the two modelling approaches using the fixed and the adaptive-mesh strategies yield almost identical 
results for the power coefficients. This re-affirms our conclusions from the mesh convergence study which states that the turbine predictions are not dependent on the mesh strategy but rather 
the edge-length used at the location of the rotor. As far as the wake field is concerned, although there is lack of data for comparison, the accuracy of the power coefficients 
predictions suggests that the model is able capture the magnitude of the wake-deficits along each row, at least in a time-averaged sense. In addition, mesh adaptivity was found to be a 
favourable choice leading to a reduction in the overall computational cost while maintaining the same accuracy. Moreover, by changing the adaptation period  $T_{\textnormal{adapt}}$, we were 
able to further reduce the overall computational cost without compromising the model's accuracy. An additional reduction of the computational cost was also observed when the adaptive 
simulations were initialised on a smaller number of CPUs, with this number increased in response to the spin--up of the computational mesh. The adaptation period was considered to be constant
during each individual simulation. Again, this is not necessarily an optimal approach, since the frequency of the adaptations should always be based on the state of the dynamics. However, such 
an approach requires relatively complex error measure designs which are beyond the scope of the present work. For an effective variable-frequency mesh-adaptive approach a 
``goal-based'' a posteriori error measure would be good to consider. This seeks to generate the optimal mesh at every instance purely for maximising the accuracy 
in a user-defined ``goal'' (e.g. power or thrust), and thus resolution is not wasted at locations and times where it does not contribute to this goal.

Regarding the limitations of the present approach, we should begin by discussing the validity of our choice to use a uniform inlet velocity with slip conditions on both the bottom and the 
top of the domain instead of a log-law profile and a rough wall model at the bottom. Such a simplification of the flow conditions were found to affect the ability of the model to predict the power output in the 
large scale simulations, particularly for the second turbine of each row. Future formulations of the mesh--adaptive solver should consider a boundary layer, although extra care should be 
taken to avoid extensive refinement near the bottom of the domain via for example using a vertically variant mesh gradation technique. In addition, by adopting a uRANS framework, many of the higher frequency interactions between the flow and the turbines were ignored and thus limiting the information that can be extracted from the simulations. To capture such effects, turbulence--resolving simulations need to be undertaken. Creech et al. \cite{CreechEtAl2015} studied the Lillgrund offshore wind farm using the 
Fluidity solver (although using different discretisation options) with an LES formulation and mesh adaptivity and found good agreement with SCADA data. However, their study did not examine the 
efficiency of mesh-adaptivity, e.g. through comparisons with fixed-mesh simulations. Inherently, an LES solver which employs mesh-adaptivity techniques will impose additional constraints on the 
frequency of mesh adaptations and the elements' aspect ratios, while a far smaller edge lengths and time step will be required. All these additional factors put the efficiency of coupling mesh 
adaptivity with an LES solver and conducting wind farm simulations in question and therefore additional studies will be needed to assess its applicability. On the other hand, coupling a uRANS 
solver with mesh adaptivity for wind energy problems may be seen as more appropriate, particularly for regional scale simulations that focus for example on the interaction of 
adjacent wind farms \citep{HansenEtAl2015}, or when the wind farm simulations are used within an adjoint--based optimisation algorithm \citep{KingEtAl2017}.

Based on the above, the mesh--adaptive uRANS framework was shown to be an appropriate and efficient tool for large scale wind farm simulations. The observed computational efficiency 
of the present simulations suggests that the same approach may be applied to more complex configurations such as those of onshore wind turbines placed over an uneven terrain and that mesh--adaptivity could be used to solve for the ensemble--averaged flow quantities (mean velocity, TKE) while requiring substantially fewer CPU hours.
\section*{Acknowledgements}
The authors would like to thank Dr James Percival for advice given on this work, and acknowledge funding from Imperial College London's Energy Futures Lab and
the EPSRC (grant numbers EP/R007470/1 and EP/L000407/1). In addition, several insightful comments and meaningful suggestions provided by the anonymous reviewers helped us 
improve the quality of the presentation of the results. Support for parallel computations was provided by Imperial College's Research Computing Service.
\bibliographystyle{wileyNJD-AMA}
\bibliography{WE2018}
\end{document}